\PassOptionsToPackage{table}{xcolor}
\PassOptionsToPackage{hyphens}{url}

\documentclass[conference,compsoc]{IEEEtran}
\ifCLASSOPTIONcompsoc
  \usepackage[nocompress]{cite}
\else
  \usepackage{cite}
\fi
\ifCLASSINFOpdf
\else
\fi
\usepackage{tikz}
\usepackage{amsmath}
\usepackage{enumitem}
\usepackage{enumerate}
\usepackage{wasysym}
\usepackage{subcaption}
\usepackage{xcolor}
\usepackage{cite}
\usepackage{booktabs}
\usepackage{makecell}
\usepackage{hyperref}
\usepackage{adjustbox}
\usepackage{todonotes}
\usepackage{tcolorbox}
\usepackage{tikz}
\usepackage[framemethod=TikZ]{mdframed}
\usepackage{wrapfig}

\usepackage[numbers]{natbib}

\usepackage{xcolor}              % load once (remove if already loaded)
\definecolor{policyblue}{HTML}{003366}
\definecolor{darkgreen}{RGB}{0,100,0}

\usepackage{tcolorbox}           % no special options; keeps it compatible
\tcbset{
  policybox/.style={
    colback=white,
    colframe=policyblue,
    boxrule=0.8pt,
    arc=2mm,                                 % rounded corners
    left=6pt,right=6pt,top=4pt,bottom=4pt,
    before skip=6pt, after skip=6pt
  }
}

\usepackage{booktabs,multirow,array,colortbl,xcolor}
\usepackage{tikz}

\definecolor{grpRC}{HTML}{D9E8F6}   % Rights & Controls
\definecolor{grpST}{HTML}{F6E1D5}   % Sharing & Transfer
\definecolor{grpMD}{HTML}{DDEAD8}   % Model Development
\definecolor{grpGM}{HTML}{CFE4F1}   % Governance & Meta
\usepackage{graphicx}   % for \scalebox (optional)
\usepackage{pifont}
\usepackage{xcolor}
\newcommand{\newstar}[1][1]{{\scalebox{#1}{\ding{72}}}}
\newcommand{\capbg}[2]{%
  \begingroup\setlength{\fboxsep}{1pt}% padding around text
  \colorbox{#1}{\strut #2}%
  \endgroup
}

\def\BibTeX{{\rm B\kern-.05em{\sc i\kern-.025em b}\kern-.08em
    T\kern-.1667em\lower.7ex\hbox{E}\kern-.125emX}}

\hypersetup{
    linkcolor=blue,
}

\newcommand{\icon}[2][]{%
  \raisebox{-0.2ex}{\includegraphics[height=1.05em,#1]{#2}}%
}

\begin{document}

\title{A Longitudinal Measurement of Privacy Policy Evolution for Large Language Models}

% \author{Anonymous Author(s)}


% version 1

\author{\IEEEauthorblockN{Zhen Tao$^1$, Shidong Pan$^{2,3}$, Zhenchang Xing$^4$, Emily Black$^3$, Talia Gillis$^2$, Chunyang Chen$^1$}
\IEEEauthorblockA{$^1$Technical University of Munich, $^2$Columbia University, $^3$New York University, $^4$CSIRO’s Data61}
% \and
% \IEEEauthorblockN{Shidong Pan}
% \IEEEauthorblockA{New York University \& Columbia University}
% \and
% \IEEEauthorblockN{Zhenchang Xing}
% \IEEEauthorblockA{CSIRO’s Data61}
% \and
% \IEEEauthorblockN{Emily Black}
% \IEEEauthorblockA{New York University}
% \and
% \IEEEauthorblockN{Talia Gillis}
% \IEEEauthorblockA{Columbia University}
% \and
% \IEEEauthorblockN{Chunyang Chen}
% \IEEEauthorblockA{Technical University of Munich}
}

\maketitle

\begin{abstract}
% Pre-trained large language models (LLMs) are rapidly deployed into various applications in everyday life 
Large language model (LLM) services have been rapidly integrated into people's daily lives as chatbots and agentic systems.
They are nourished by collecting rich streams of data, raising privacy concerns around excessive collection of sensitive personal information. 
Privacy policies are the fundamental mechanism for informing users about data practices in modern information privacy paradigm. 
Although traditional web and mobile policies are well studied, the privacy policies of LLM providers, their LLM-specific content, and their evolution over time remain largely underexplored. 
In this paper, we present the first longitudinal empirical study of privacy policies for mainstream LLM providers worldwide. We curate a chronological dataset of 74 historical privacy policies and 115 supplemental privacy documents from 11 LLM providers across 5 countries up to August 2025, and extract over 3,000 sentence-level edits between consecutive policy versions.
% \sd{I think one sentence to summarize methods, and three sentences for each RQ/highlighted results?}
% We first compare LLM policies with established baselines for general software, showing that they are substantially longer, require college-level reading ability, and remain highly vague. 
% We then propose a taxonomy tailored to LLM privacy policies and reveal patterns in how providers disclose LLM-specific practices as well as regional disparities. 
% Lastly, we annotate policy edits and align them with a timeline of LLM ecosystem events, finding that first-party data collection, product releases and regulatory actions are key drivers for edits. Our findings shed light on the status quo of LLM privacy policies and how they evolve.
We compare LLM privacy policies to those of other software formats, propose a taxonomy tailored to LLM privacy policies, annotate policy edits and align them with a timeline of key LLM ecosystem events. 
Results show they are substantially longer, demand college-level reading ability, and remain highly vague. Our taxonomy analysis reveals patterns in how providers disclose LLM-specific practices and highlights regional disparities in coverage. Policy edits are concentrated in first-party data collection and international/specific-audience sections, and that product releases and regulatory actions are the primary drivers, shedding light on the status quo and the evolution of LLM privacy policies.
% We first compare LLM policies with established baselines for general software, showing that they are substantially longer, require college-level reading ability, and remain highly vague. We then propose a taxonomy tailored to LLM privacy policies and reveal patterns in how providers disclose LLM-specific practices as well as regional disparities. Lastly, we annotate policy edits and align them with a timeline of LLM ecosystem events, finding that first-party data collection, product releases and regulatory actions are key drivers for edits. Our findings shed light on the status quo of LLM privacy policies and how they evolve.
\end{abstract}


% \fi
\IEEEpeerreviewmaketitle

\section{Introduction}
\label{sec_intro}

Large Language Models (LLMs) and their applications, e.g., ChatGPT~\cite{ChatGPT}, offer users convenience, creativity, and new digital services at scale. 
At the same time, providers of LLMs such as OpenAI collect and process substantial amounts of personal data through prompts, uploaded files and other user interaction data, raising significant concerns regarding users' privacy~\cite{ali2025understanding, kwesi2025exploring, malki2025hoovered, king2025user}. 
More recent multimodal features and agentic systems have extended the scope to include audiovisual information and third-party tool-use traces~\cite{GPT4o, ChatGPTplugins}. Unlike traditional software such as Android apps with explicit permission systems, LLM interactions are often open-ended, making it harder to identify which data types are collected and how they are processed, shared and even reproduced. These characteristics introduce unique privacy risks, such as the unintended exposure of
sensitive information~\cite{gupta2023chatgpt, kibriya2024privacy, li2023privacy}. For instance, in response to data security risks, Samsung restricted employee use of ChatGPT after a leak incident of sensitive company code.\footnote{\href{https://www.forbes.com/sites/siladityaray/2023/05/02/samsung-bans-chatgpt-and-other-chatbots-for-employees-after-sensitive-code-leak/}{forbes.com/samsung-bans-chatgpt-after-sensitive-code-leak/}} 

In response to privacy challenges in the era of LLMs, privacy and AI governance offer regulatory remedies. Privacy laws such as European General Data Protection Regulation (GDPR)~\cite{GDPR} and California Consumer Privacy Act (CCPA)~\cite{CCPA} set obligations on lawful bases of data processing, transparent notices and user rights that still apply to LLM providers. For instance, OpenAI's insufficient disclosure of data processing in its privacy policy was ruled by the Data Protection Authority in Italy to have violated the transparency obligations in GDPR (Art.5(1)(a), 12 and 13) and ultimately resulted in a €15 million fine. Emerging AI-specific regulations, e.g., the EU Artificial Intelligence Act (EU AI Act)~\cite{AIAct}, add obligations on transparency, documentation, and disclosures about training data and model behavior. 
These frameworks have implemented influence on how LLM providers construct their privacy policies to disclose data practices and state users' data rights. 

Privacy policies have become the most common mechanism for informing users about how their personal data will be handled~\cite{adams2020agreeing, bui2023detection, caramujo2015analyzing, perez2018review, harkous2018polisis, kemp2020concealed, tao2025privacy, wang2025big}. 
Meanwhile, researchers have criticized privacy policies for being long, complex, and ambiguous~\cite{singh2011evaluating, ermakova2015readability, fabian2017large, adhikari2023evolution, amos2021privacy, wagner2023privacy}. 
In parallel, users continue to exhibit ``digital resignation''~\cite{draper2019corporate}, where they express concerns about privacy but feel powerless to understand privacy policies and act on privacy choices. A recent study~\cite{malki2025hoovered} in the LLM context reinforces this gap, reporting that only 7\% users opted out of their data being used to train models in a survey of 211 conversational agent users, and many participants held mismatched expectations about opting out of training or requesting deletion.
However, most prior studies analyze policies of traditional software, such as websites~\cite{wilson2016creation, belcheva2023understanding} and mobile apps~\cite{verderame2020reliability, cui2023poligraph, qiu2023calpric}, while LLM privacy policies remain underexplored. 
Therefore, it is essential to reveal how LLM providers’ policies compare to traditional generations, how they evolve in response to events such as product launches and regulatory milestones, and which LLM-specific categories drive changes. Thus, this study explores the following research questions:

\begin{itemize} [leftmargin=*, noitemsep, topsep=3pt]
\item \textbf{RQ1} How do privacy policies for LLMs differ from those of traditional software?
\item \textbf{RQ2} What are the differences between privacy policies across LLM providers and jurisdictions?
% \item How have privacy policies evolved over time in response to events, such as technological advancements and regulatory shifts?
\item \textbf{RQ3} How do privacy policies evolve over time, and how do key events in the LLM ecosystem have an impact on this evolution?
\end{itemize}

% \chen{Some paragraph is too long, and try to split them into two to make overall format consistent.}
Existing studies touch parts of the problem by analyzing LLM app privacy policies~\cite{wu2024depth}, conducting case studies~\cite{naghiyev2024chatgpt, wu2024unveiling}, and relatively small-scale document coding~\cite{king2025user} on LLM policies, but rarely cover both low-level document analysis and high-level regulatory impact in a longitudinal manner.
% but rarely summarize new patterns in a larger scale or connect time-stamped edits to concrete external events. 
% \sd{Our strengths are we cover both low-level document analysis, high-level regulatory impact, and the chronological factor, all together.}
Most established analysis approaches typically rely on classic taxonomies, e.g., OPP-115~\cite{wilson2016creation}, which may under-represent LLM-specific characteristics such as training data handling and third-party plugin integrations. 
% To address these gaps, we conduct an empirical, longitudinal study of 11 mainstream LLM providers’ privacy policies. 
To address these gaps, we conduct an empirical study of 11 mainstream LLM providers’ privacy documents. 
We collect a corpus of 74 historical versions of privacy policies and 115 supplemental documents up to August 2025. 
We find that they are around 53.6\% longer than the traditional software policies baseline~~\cite{wagner2023privacy,amos2021privacy}.
% We measure their length, readability, and vagueness, contrasting LLM policies with traditional baselines. \zhen{We found that...} 
We introduce a new taxonomy tailored to LLM privacy policies based on the OPP-115 taxonomy~\cite{wilson2016creation} and annotate the latest version of policies to identify patterns.

Results show that LLM privacy policy has expanded in multiple dimensions, including new data types, first-party and third-party data interactions, model training, and user rights. 
To identify where and how LLM policies have evolved over time, we apply a lightweight heuristic to extract 3,463 sentence-level edits between consecutive policy versions. 
We then manually annotate the categories of policy edits based on our taxonomy. 
%\chen{Do we have that?}
In addition, we align policy updates with a timeline of key events that impact the LLM ecosystem to understand whether and how events may influence the updates of LLM privacy policies. 
LLM providers often update privacy policies alongside product launches. Regulatory pressure also influences policy changes, but often with a lag, as responding to requests from data protection authorities may take time. We find providers may de-specify regulations from explicit references to broader phrases. We identify a copy–paste–modify reuse of clauses across providers. These findings map where LLM policies are growing and how they evolve.

\textbf{Contributions.} To our knowledge, this is the first empirical study of privacy policies from LLM providers worldwide and longitudinal measurement of their evolution. Our observations and insights will benefit stakeholders including LLM providers, downstream AI developers, and privacy regulators. In summary, we make the following contributions:
\begin{itemize} [leftmargin=*, noitemsep, topsep=3pt]
\item We conduct the first empirical study that reveals the new pattern of LLM privacy policies worldwide. We introduce a new privacy policy taxonomy tailored to LLMs, offering novel insights into LLM data practice disclosures.
\item We curate a dataset of 74 historical versions of privacy policies and 115 supplemental documents up to August 2025 for mainstream LLMs, spanning 11 companies from 5 countries. In addition, our dataset includes 3,463 sentence-level edits between consecutive policy versions. 
\item We conduct a longitudinal measurement to understand how LLM privacy policies have evolved and how key events such as technological advancements and regulatory shifts in the LLM ecosystem may influence this evolution.
% \item We conduct a chronological analysis to track the evolution of LLM privacy policies and reveal how key events, such as technological advancements, regulatory shifts, and public scrutiny, in the LLM ecosystem may influence the updates of LLM privacy policies.
\end{itemize}

% We argue that LLMs introduce greater complexity and unique challenges compared to traditional software. 
% \begin{itemize}
%     \item The uncertainty of software behavior makes static/dynamic analysis more challenging.
%     \item The data collection methods are different to traditional software. Android app has a permission system to track data practices. AI (e.g., ChatGPT) might take user prompts as the input, and it is challenging to accurately track user data types from longer and longer prompts nowadays.
%     \item The AI regulations are different from traditional privacy law/regulations. 1) AI regulations are emerging and dynamic. 2) AI regulations are on specific techniques details, making it easier to implement. 3) how to combine AI regulations and privacy laws/regulations. Opportunities and challenges co-exist.
% \end{itemize}

\section{Background}
\label{sec_background}

% In this section, we present the history of LLMs and the accompanying privacy challenges, regulations, and privacy policy studies.

% \chen{Should the related work part be in the later paper? Not sure the common norm in security domain.}
% \sd{We need to separate the background and related work. For related work, we can have other large-scale/chronological PP studies. }

% \subsection{The History of LLMs and Privacy Challenges}
% AI chatbots ranked by data they collect by Surfshark~\cite{chatbotsrank}. 
% \zhen{around 3 generations for LLMs. 0: bert; 1: gpt3; 2: multi-modal..., reasoning, R1..., related to privacy policy, data collection, reasoning location...3: agent...perform tasks...related to privacy, trade-offs privacy and utility, more data collection...}
\subsection{Evolution and Accompanying Privacy Risks}
LLM development has unfolded in several broad stages, each expanding technical capability and introducing unprecedented privacy risks. 
Early models centered on contextual encoders such as BERT~\cite{devlin2019bert}, which performed downstream NLP tasks through fine-tuning. The privacy exposure was relatively limited to training corpus leakage~\cite{carlini2021extracting, huang2022large, mireshghallah2022quantifying, carlini2022membership, chen2020gan}. The generative turn with GPT-3~\cite{brown2020language} demonstrated few-shot abilities and catalyzed consumer chat products such as ChatGPT~\cite{ChatGPT}. 
This shift has led to the extensive collection and use of user interaction data~\cite{king2025user}. Multimodal and reasoning models then merged, including GPT-4o~\cite{GPT4o} and DeepSeek-R1~\cite{guo2025deepseek}. This stage expands data collection beyond text to audio, images, and long documents~\cite{meeus2024did, hu2025membership}. 
At the current stage, systems increasingly behave as agents, with features such as tool calling and web browsing. 
Such mode creates privacy–utility trade-offs, where richer capability often entails more integration points and data sharing~\cite{iqbal2024llm}. LLM agent privacy risks have caused concerns~\cite{gan2024navigating, wang2025unveiling, he2024emerged, dong2023philosopher}.
% and have been widely studied
Over time, privacy has become a moving target that co-evolves with the evolution of LLM functionality.

\subsection{Privacy Regulations}
% \zhen{regulation name, logic behind the arising of regulations, concerns. Tech part, covers what}
% China's Basic Security Requirements for Generative Artificial Intelligence Services: \url{https://www.tc260.org.cn/upload/2024-03-01/1709282398070082466.pdf}
%\sd{@Talia, can you help have a look? Also can you help convert those law/legal references into the footnote law style? }
Privacy and AI governance have evolved worldwide to answer concerns about personal data protection and the risks introduced by LLMs. 
In the EU, the GDPR~\cite{GDPR} from 2016 provides the cornerstone of data governance, imposing principles on personal data handling and requiring that data be handled ``\textit{lawfully, fairly and in a transparent manner}''(Art. 5(1)(a)). 
A GDPR enforcement case~\cite{gdprhubItaly} illustrates that data-protection principles are being applied in  LLM contexts. In this case, Italy’s data protection authority found that OpenAI’s ChatGPT processed users' personal data  without adequate legal basis, and also failed to satisfy transparency and information-obligations, resulting in a €15 million fine.
Furthermore, the EU AI Act~\cite{AIAct} from 2024 introduces a risk-based regulatory framework for AI systems, including general-purpose AI (GPAI) models, highlighting transparency and safety obligations. It establishes obligations for providers of AI systems regarding risk management, data governance and technical documentation. Similarly, China’s Basic Security Requirements for Generative Artificial Intelligence Services\footnote{\href{https://cset.georgetown.edu/publication/china-safety-requirements-for-generative-ai-final/}{https://cset.georgetown.edu/publication/china-safety-requirements-for-generative-ai-final/}} set granular safety controls and specify obligations on transparency, requiring information such as ``personal information collected and its uses in the service'' to be disclosed in ``easily viewed locations''.
In the United States, Executive Order 14110~\cite{EXECUTIVEORDER14110} from 2023 launched a comprehensive federal initiative on AI governance, directing agencies to develop ``new standards for AI safety and security'' and  to protect Americans' privacy. The Order's emphasis on data protection and transparency affects both traditional data practices (e.g., lawful bases, user rights) and AI-specific obligations (e.g., model transparency, data governance) reflected in LLM privacy policies. Similar developments are emerging in other jurisdictions, such as UK~\cite{ICOAIGuidance}, Canada~\cite{AIDA} and Brazil~\cite{GAIBrazil}.

\begin{table*}[t]
\centering
\caption{Our chronological dataset of historical versions of privacy policies and privacy-related supplementary documents for LLMs. ``Count'' refers to the number of unique historical versions of privacy policies. The last three columns refer to the the number of distinct supplementary documents at the Last ``Effective Date''. we did not include Meta and Google's main privacy policies in the dataset based on our criteria and considerations elaborated in Section~\ref{subsec_pp_collection}. Instead, we collect their AI-specific supplemental documents. In the rest of the paper, we alternatively use (P1) to (P11) to indicate the LLM provider as listed in the table.
}
\label{tab_LLM_provider}
\resizebox{.99\textwidth}{!}{%
\begin{tabular}{l|l|c|c|c|c|c|c|c}
\toprule
\textbf{LLM Provider} & \textbf{Jurisdiction} & \textbf{Count} & \textbf{Avg. \#Words} & \textbf{First ``Effective Date''} & \textbf{Last ``Effective Date''}  & \textbf{\#Addendum}  & \textbf{\#Help Articles} & \textbf{\#Add'l. Terms}\\
\midrule
\midrule
\raisebox{-0.0\totalheight}{
1~
\begin{minipage}{.04\textwidth}
  \includegraphics[width=0.85\linewidth]{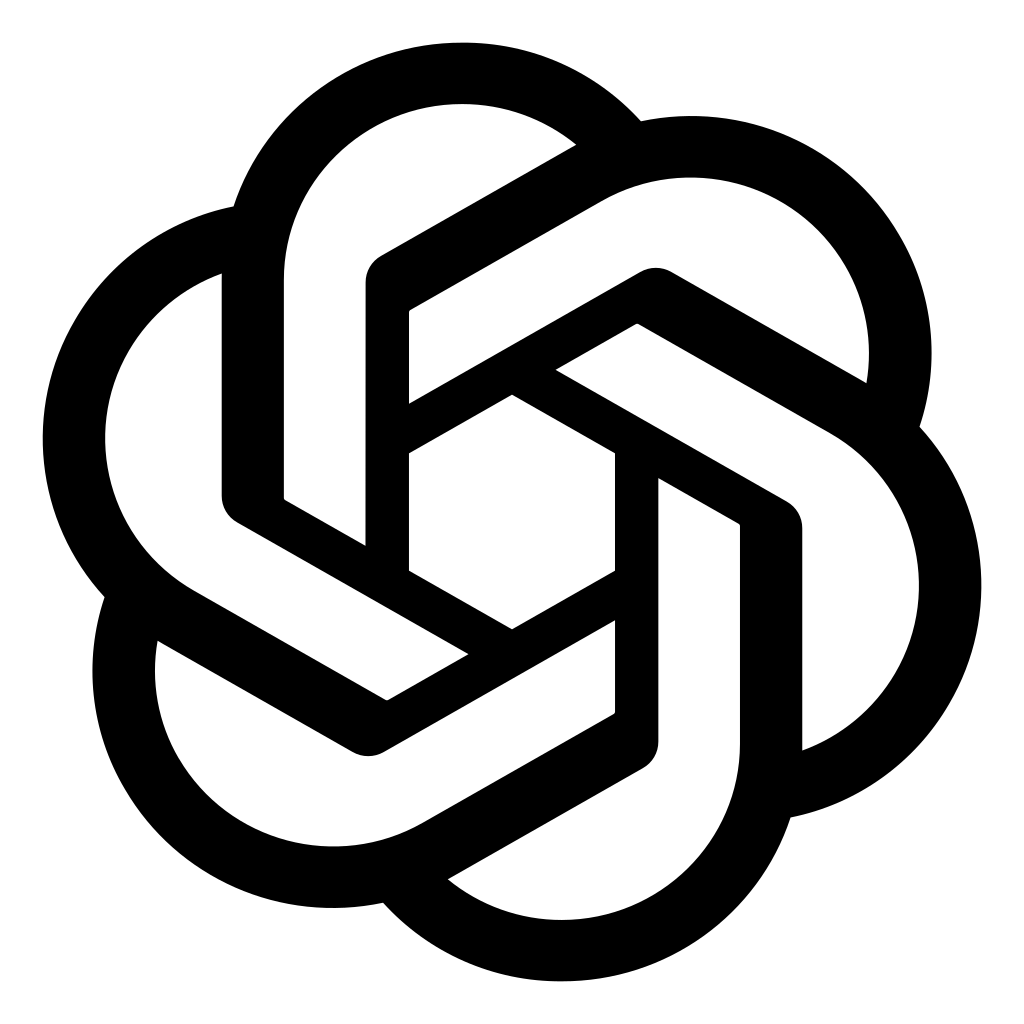}
\end{minipage}
OpenAI
} & United States & 22 & 2,797 & Feb, 2021 & Jun, 2025 & 1 & 2 & 2 
\\ \hline
\raisebox{-0.0\totalheight}{
2~
\begin{minipage}{.04\textwidth}
  \includegraphics[width=0.85\linewidth]{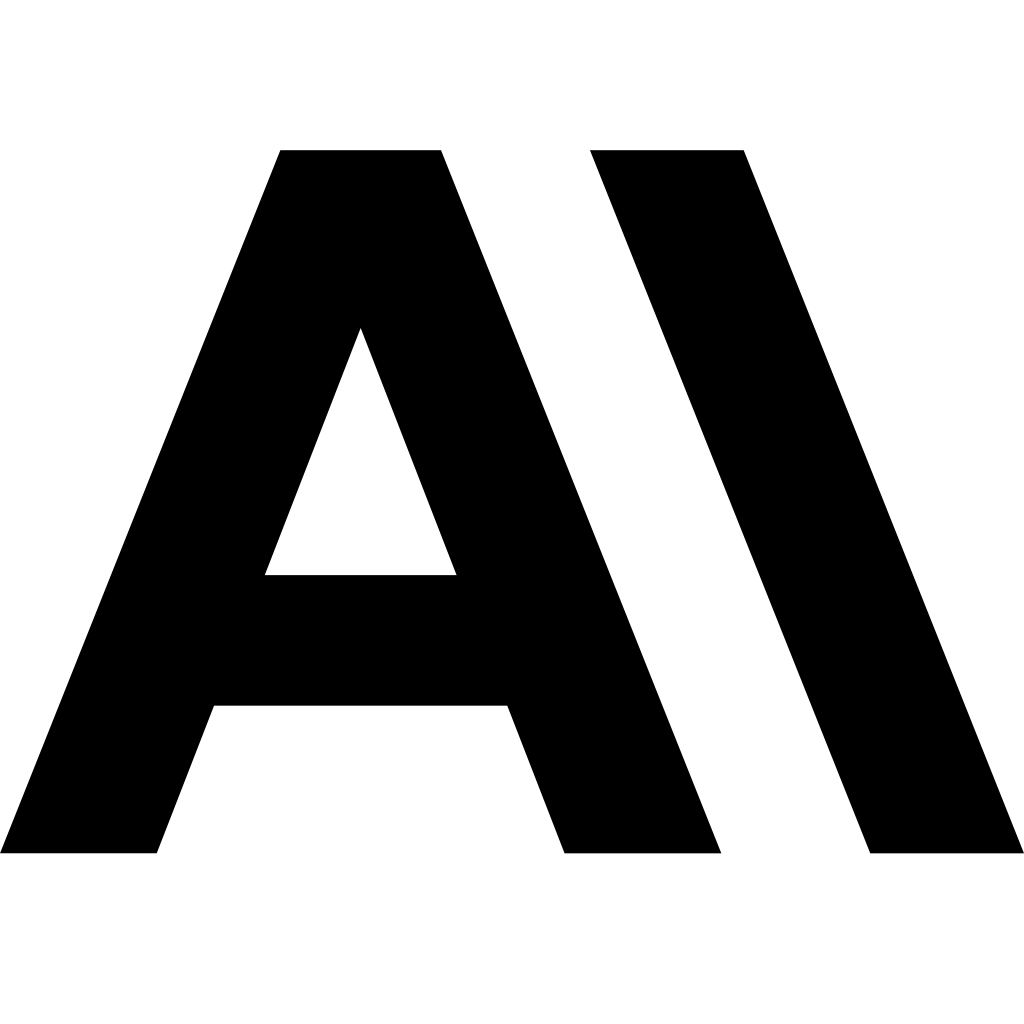}
\end{minipage}
Anthropic
} & United States & 13 & 3,768 & Feb, 2023 & May, 2025 & 1 & 24 & 2 
\\ \hline
\raisebox{-0.0\totalheight}{
3~
\begin{minipage}{.04\textwidth}
  \includegraphics[width=0.85\linewidth]{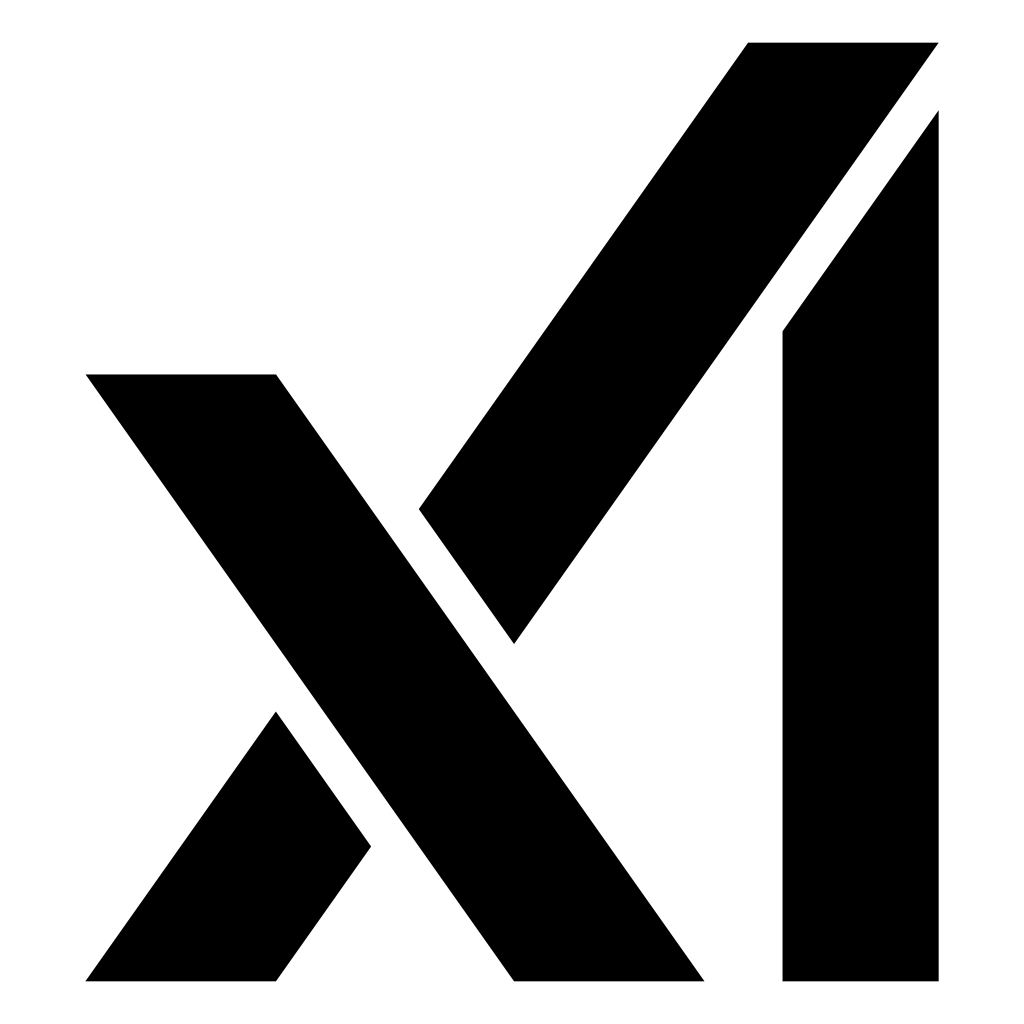}
\end{minipage}
xAI
} & United States & 9 & 5,646 & Nov, 2023 & Jul, 2025 & 1 & 2 & 1 
\\  \hline
\raisebox{-0.0\totalheight}{
4~
\begin{minipage}{.04\textwidth}
  \includegraphics[width=0.85\linewidth]{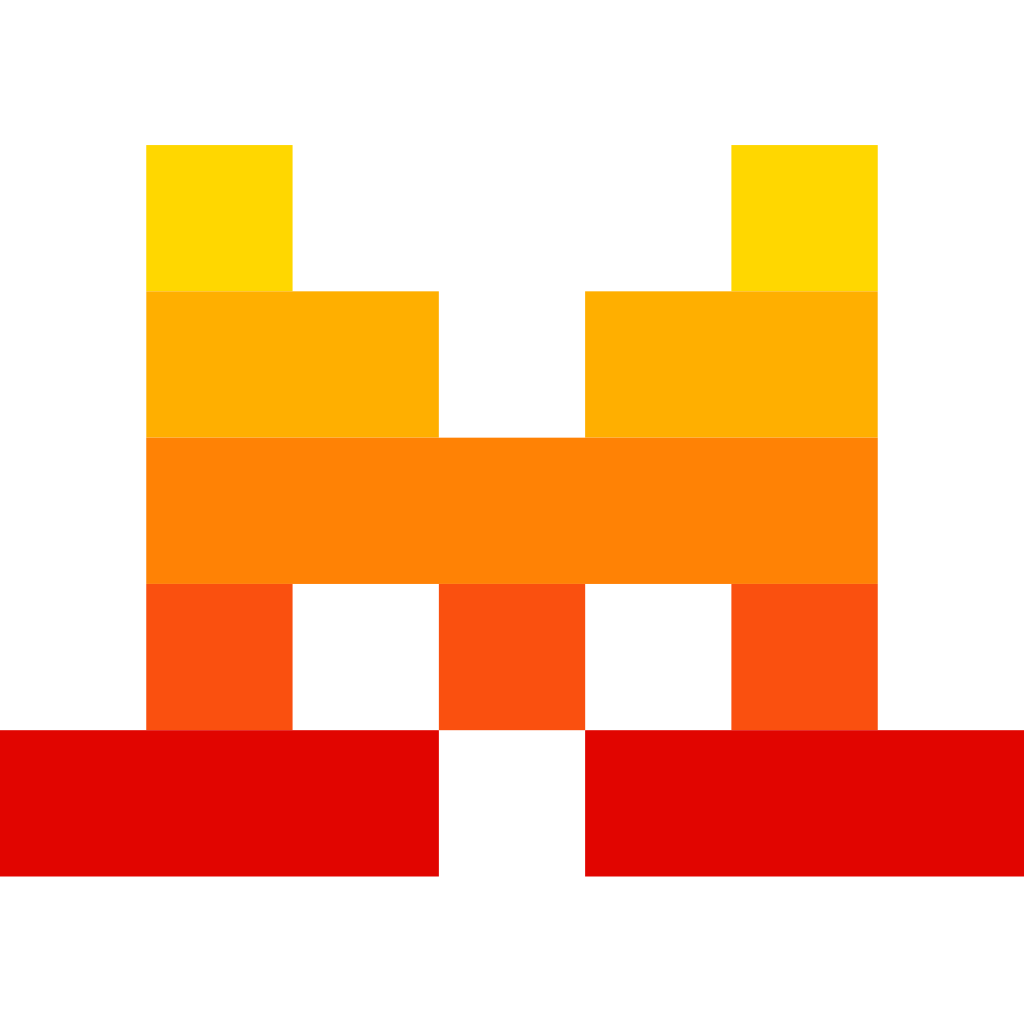}
\end{minipage}
Mistral 
} & France & 12 & 2,462 & Sep, 2023 & May, 2025 & 0 & 0 & 5 
\\ \hline
\raisebox{-0.0\totalheight}{
5~
\begin{minipage}{.04\textwidth}
  \includegraphics[width=0.85\linewidth]{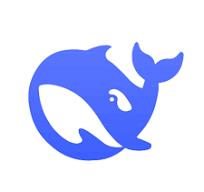}
\end{minipage}
DeepSeek
} & China & 7 & 3,625 & Nov, 2023 & Jul, 2025 & 0 & 0 & 1 
\\ \hline
\raisebox{-0.0\totalheight}{
6~
\begin{minipage}{.04\textwidth}
  \includegraphics[width=0.85\linewidth]{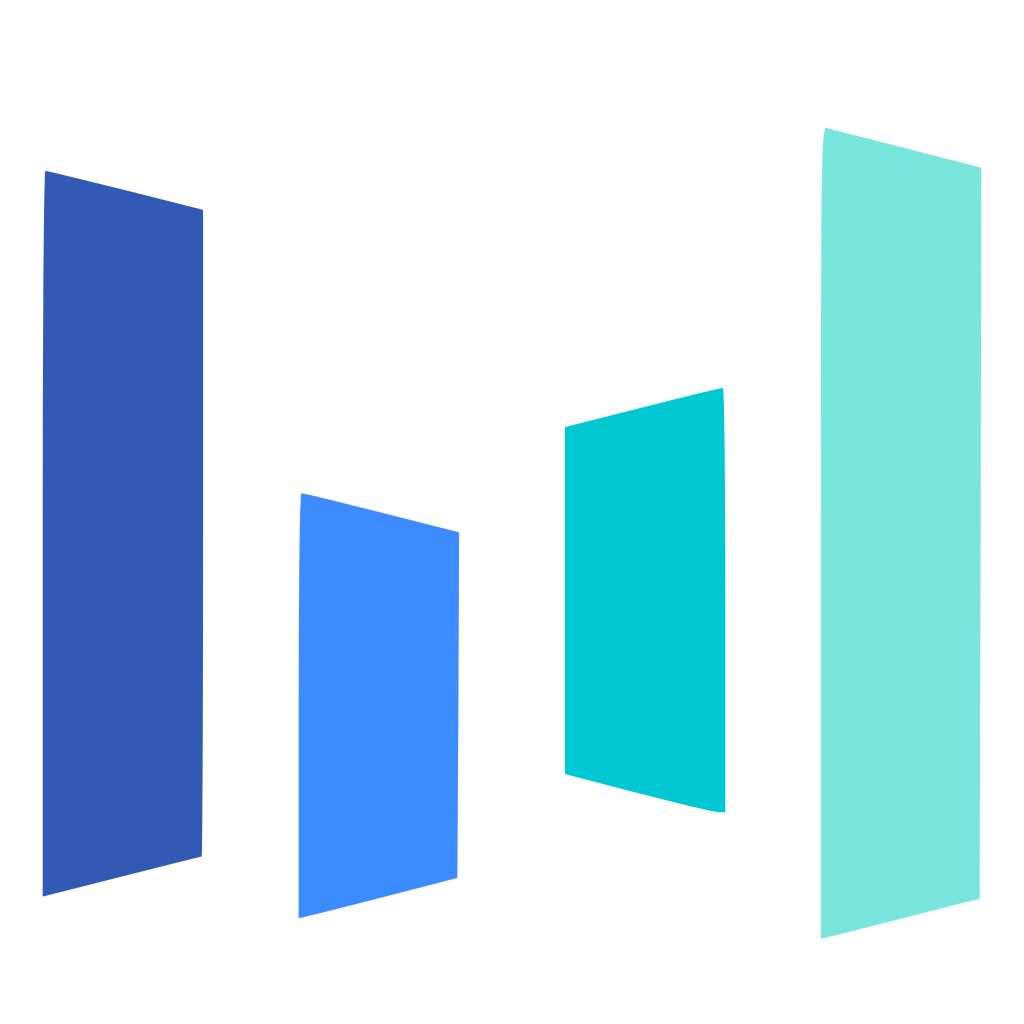}
\end{minipage}
ByteDance
} & China & 1 & 2,113 & May, 2024 & May, 2024 & 0 & 0 & 0 
\\ \hline
\raisebox{-0.0\totalheight}{
7~
\begin{minipage}{.04\textwidth}
  \includegraphics[width=0.85\linewidth]{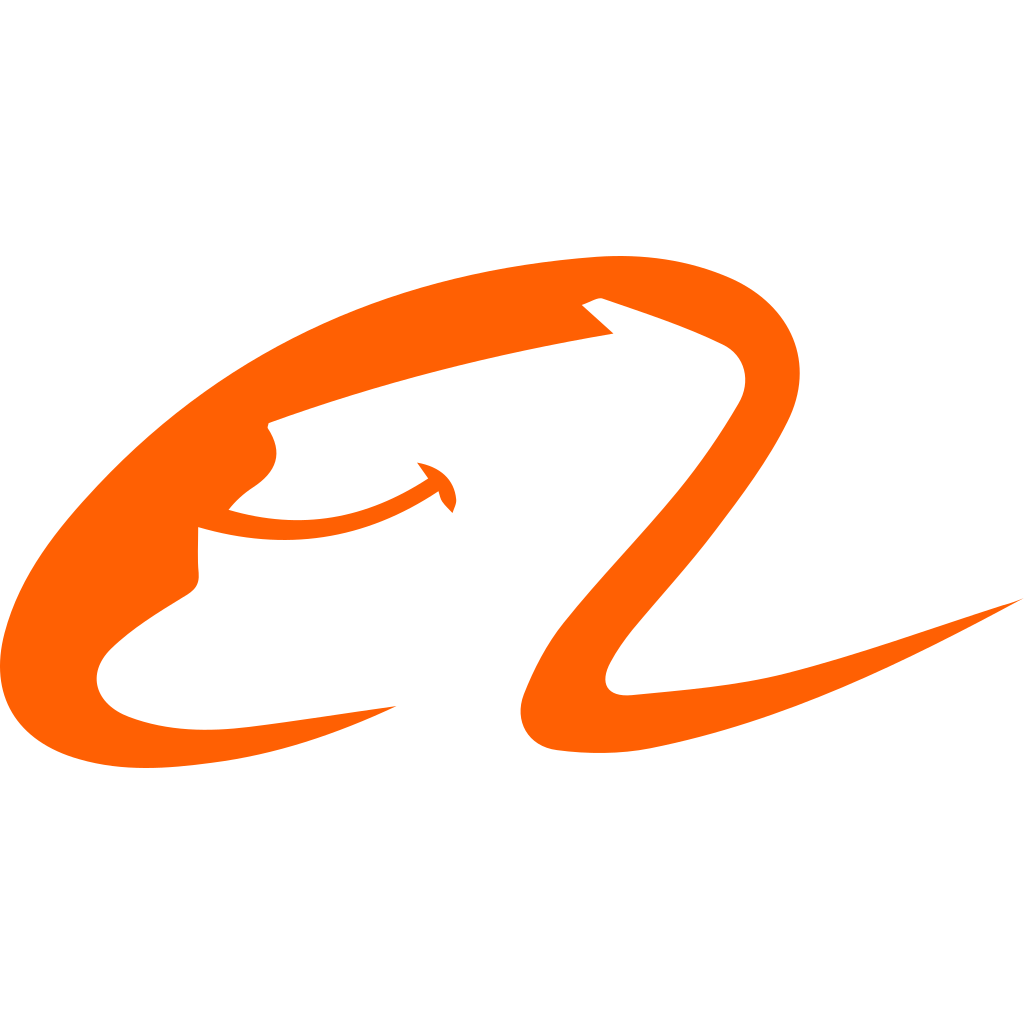}
\end{minipage}
Alibaba
} & China & 3 & 3,447 & Jan, 2025 & Apr, 2025 & 0 & 0 & 0 
\\ \hline
\raisebox{-0.0\totalheight}{
8~
\begin{minipage}{.04\textwidth}
  \includegraphics[width=0.85\linewidth]{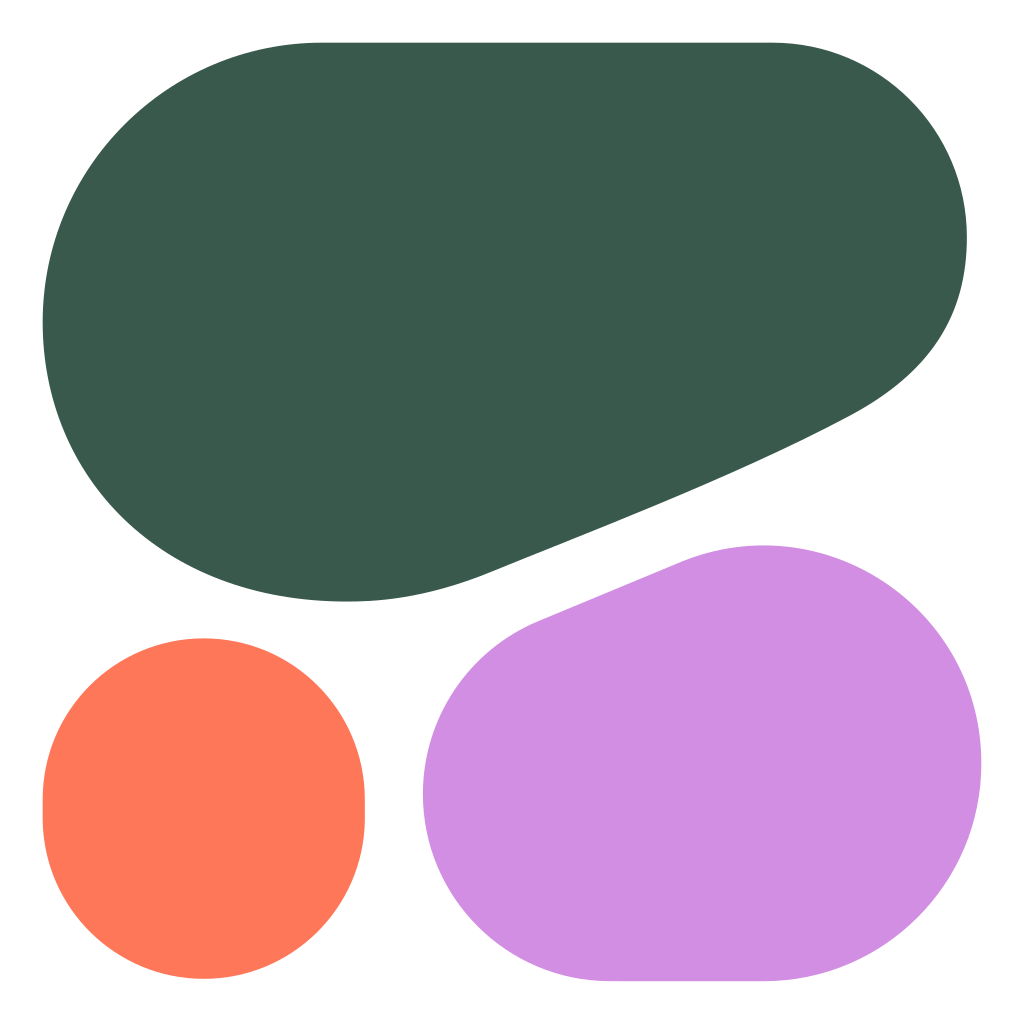}
\end{minipage}
Cohere
} & Canada & 5 & 2,166 & Apr, 2021 & Sep, 2024 & 0 & 0 & 1 
\\ \hline
\raisebox{-0.0\totalheight}{
9~
\begin{minipage}{.04\textwidth}
  \includegraphics[width=0.85\linewidth]{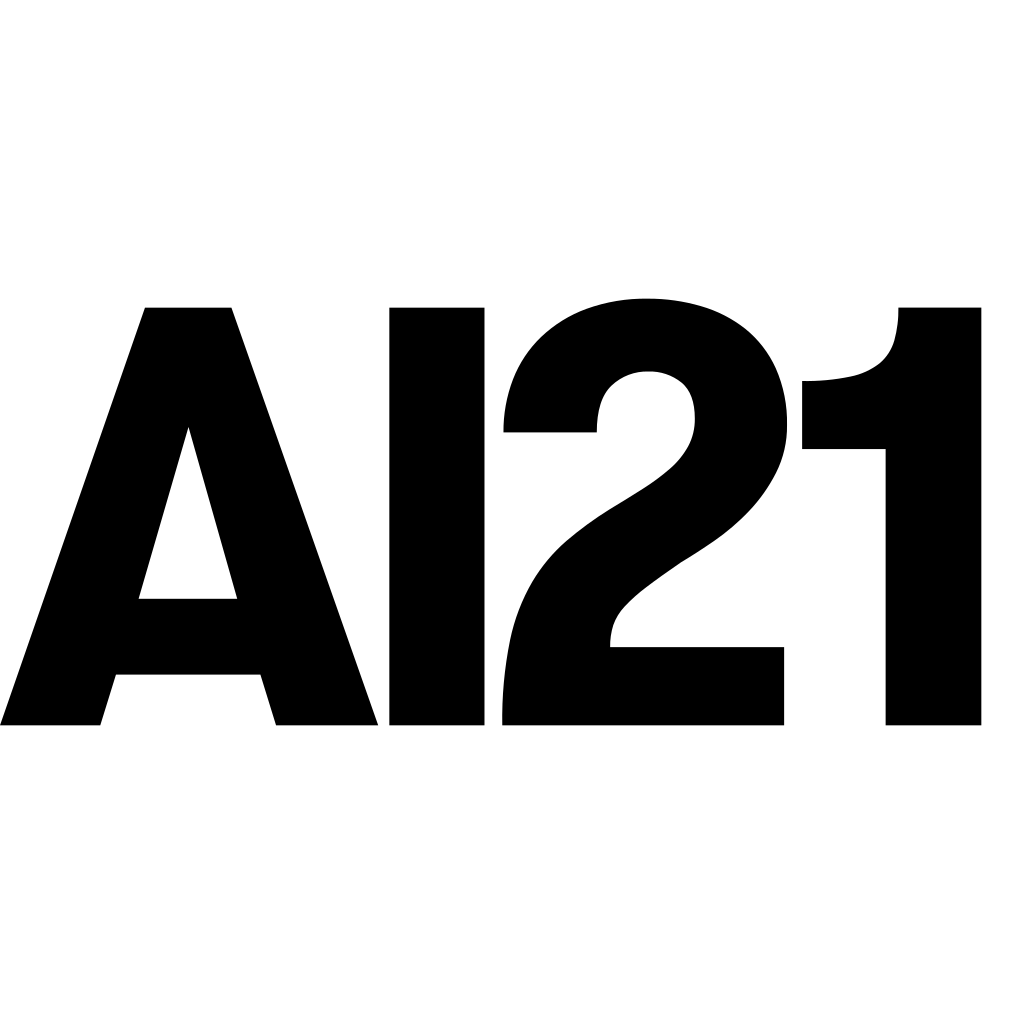}
\end{minipage}
AI21 Labs
} & Israel & 2 & 4,067 & Aug, 2021 & Jul, 2025 & 0 & 1 & 1 
\\
\hline
\hline
\rowcolor{lightgray!30}
\raisebox{-0.0\totalheight}{
10
\begin{minipage}{.04\textwidth}
  \includegraphics[width=0.85\linewidth]{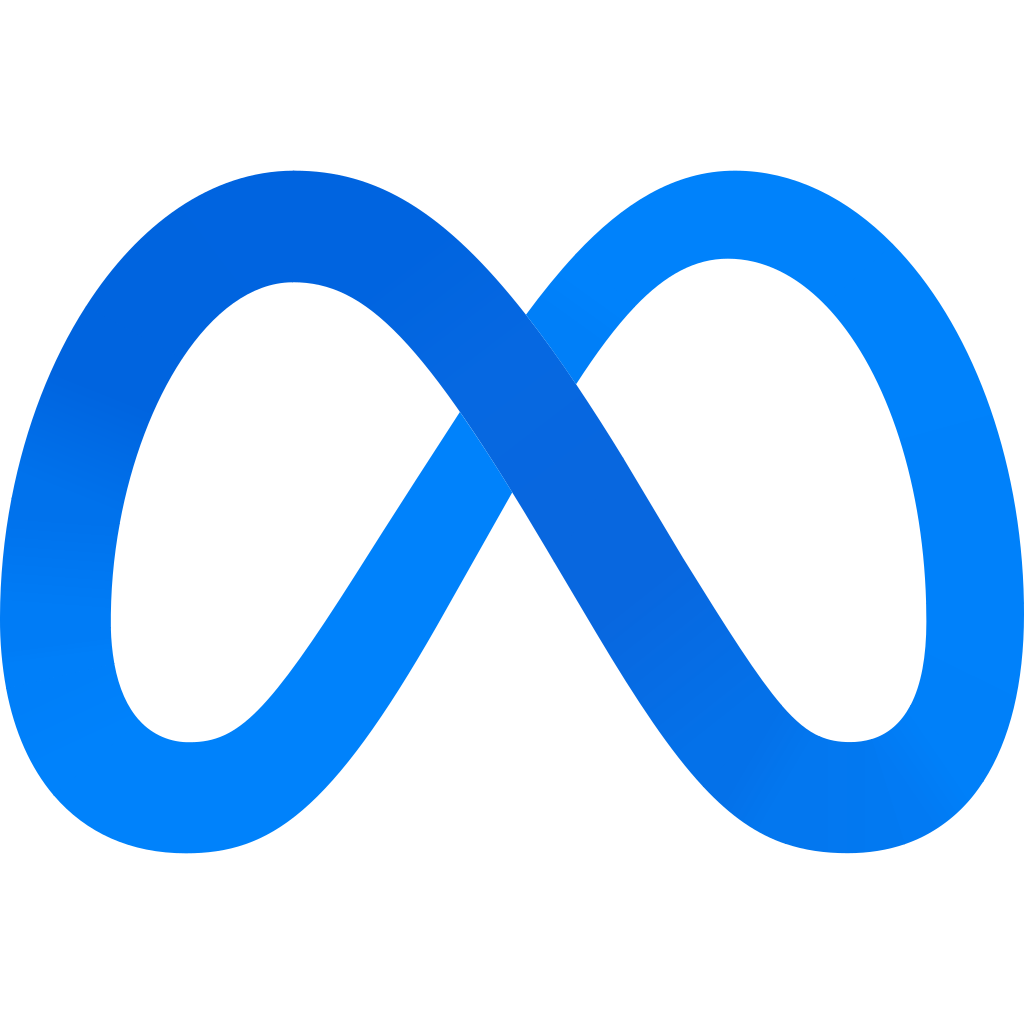}
\end{minipage}
Meta
} & United States & - & - & - & - & 0 & 0 & 4 
\\ \hline
\rowcolor{lightgray!30}
\raisebox{-0.0\totalheight}{
11
\begin{minipage}{.04\textwidth}
  \includegraphics[width=0.85\linewidth]{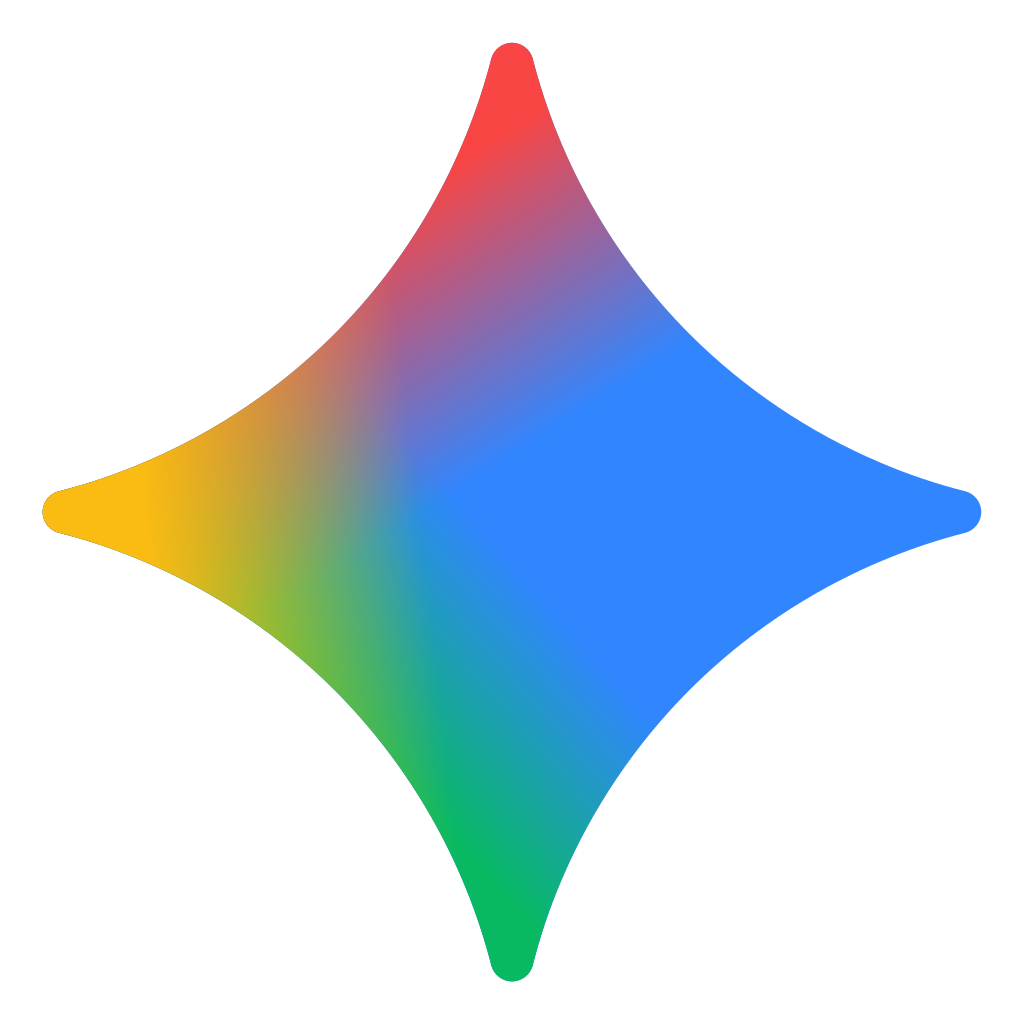}
\end{minipage}
Google
} & United States & - & - & - & - & 0 & 0 & 3 
\\

% $\textit{Q}_1$  &\makecell[l]{How concerned are you about your privacy information \\ while using mobile apps?} & 5 for very concerned, 1 for very unconcerned & 4.13 & 4 &  1.09\\
% \midrule
% $\textit{Q}_2$    & \makecell[l]{Do you read mobile app's privacy policies when \\you encountered?} &  5 for always read, 1 for never read &2.00 & 2 & 1.03 \\

\bottomrule
\end{tabular}
}%
\end{table*}
%
%
%----

\subsection{Privacy Policy Analysis}

% \sd{I dont think we need small titles for each paragraphs.}
% \textbf{Text Feature Analysis.} 
To meet regulatory disclosure requirements, companies provide users with privacy documents, such as privacy policies, to inform users of various data practices. However, privacy policies are often lengthy, problematic and difficult to create~\cite{balebako2014improving, balebako2014privacy, li2022understanding, pan2024new, tao2025privacy}, which limits their effectiveness for informing users. The readability and vagueness of privacy policies is a long-standing issue and has attracted much research attention. Early studies in mobile apps~\cite{singh2011evaluating} and healthcare websites~\cite{ermakova2015readability} reported low readability and comprehension, highlighting dense text and user burden. A large-scale readability analysis~\cite{fabian2017large} provided empirical evidence that policies are hard to read.
% Liu et al.~\cite{liu2016modeling} modeled vagueness leveraging deep neural networks
Recent longitudinal studies~\cite{adhikari2023evolution, amos2021privacy, wagner2023privacy, belcheva2023understanding} also show readability and vagueness remain a problem and the composition of policies has grown more complex over two decades. Beyond text quality, policy length itself is a hurdle, as longer policies can hinder comprehension and decision-making~\cite{meier2020shorter}. 
However, prior studies have either been large-scale studies of the market or focused on a specific category such as health websites~\cite{ermakova2015readability}, and empirical studies on the privacy practices of mainstream LLM providers are relatively limited~\cite{king2025user}. Motivated by this gap, we examine the length, readability, and vagueness of LLM policies, revealing their unique content and evolution.

% \textbf{Readability and vagueness:} Modeling language vagueness in privacy policies using deep neural networks~\cite{liu2016modeling}. Evolution of Composition, Readability, and Structure of Privacy Policies over Two Decades~\cite{adhikari2023evolution}. Ambiguity and Generality in Natural Language Privacy Policies~\cite{hosseini2021ambiguity}. Ambiguity in privacy policies and the impact of regulation~\cite{reidenberg2016ambiguity}. Evaluating the readability of privacy policies in mobile environments~\cite{singh2011evaluating}. Readability of privacy policies of healthcare websites~\cite{ermakova2015readability}. Large-scale readability analysis~\cite{fabian2017large}. \textbf{Availability:} Web Privacy Policy Availability~\cite{srinath2023privacy}. \textbf{Policy length:} effects of privacy policy length on online privacy decision-making~\cite{meier2020shorter}. \textbf{Privacy labels:} Longitudinal Measurement of the Adoption Dynamics in Apple’s Privacy Label~\cite{balash2023poster}. 

% \textbf{Compliance Checking.}
Beyond text features, substantial research efforts have been made to examine compliance of privacy policies. Foundationally, Wilson et al.~\cite{wilson2016creation} created a corpus of 115 policies with 23k fine-grained data-practice annotations. They proposed a taxonomy for privacy policies, which is widely used in the following studies on the structure and scope of data practices~\cite{harkous2018polisis, qiu2023calpric, xie2025evaluating, bui2021automated, hosseini2021unifying, ahmad2020policyqa, srinath2020privacy}. Amos et al.~\cite{amos2021privacy} reported consistently poor transparency over two decades, highlighting the gap between regulatory expectations and actual disclosures. Linden et al.~\cite{linden2018privacy} assessed policies before and after GDPR, evidencing GDPR’s positive effect on improving inconsistent compliance. Contradiction and inconsistency detection frameworks~\cite{andow2019policylint, bui2021consistency} have been proposed and revealed issues with policy quality.

% We contribute to the compliance analysis by extending the OPP-115 taxonomy~\cite{wilson2016creation} to the LLM context and annotating the privacy policies for mainstream LLM providers, and analyzing the impact of key events, including regulatory shifts, on the evolution of these privacy policies. \chen{Really? Do we touch that? which section?} \sd{we need to move this to the methodology section}
\section{Methodology}
\label{sec_methodology}
In this section, we present how we identify our targeted mainstream LLM providers and construct the chronological dataset of historical privacy policies. We also discuss methodologies applied in our privacy policy analysis.

\subsection{Historical Privacy Policies Collection}
\label{subsec_pp_collection}
% In this section, we present how we identify our targeted mainstream LLM providers and collect the latest and historical versions of privacy policies and additional documents for these LLMs. 

\textbf{Identification of Mainstream LLM Providers.} %The mainstream LLM providers' actions have a significantly wide impact on the market, we therefore target them in this paper. 
First, we select 19 LLM providers that are most frequently mentioned in benchmark academic papers~\cite{zeng2024air, chiang2024chatbot, he2024multi, yang2025realfactbench, gao2025mcp}, AI index and market-analysis reports~\cite{bommasani2023eu-ai-act, AIindex2025, grandviewresearch}, and tech-commentary articles~\cite{techtarget, explodingtopics, splunk, directindustry}. 
As our goal is to analyze the privacy policies of providers of mainstream foundational models or services, we exclude providers whose models do not gain internationally widespread adoption and those of privacy documents are not in English at the time of collection (Aug. 2025). 
% , such as Aleph Alpha~\cite{alephalpha}, whose services are built on top of third-party base models, such as Perplexity~\cite{PPLX}, or whose privacy documents are not in English at the time of collection, such as Baidu~\cite{ERNIE}. 
We eventually narrow down to 11 LLM providers, as presented in Table~\ref{tab_LLM_provider}.

\textbf{Document Collection.} We manually browse the official websites of target LLM providers to collect their privacy policies.
In addition, we find there are supplemental documents referenced in the policy or available on provider's website, providing additional privacy-related information beyond policy content.
These supplemental documents include addendum, help articles, and other additional terms. 
For instance, OpenAI prepares a ``\textit{Data processing addendum}'' to describe details on data practices.
To cover such disclosures and collect a full list of privacy-related supplemental documents, we utilize the site map in Wayback Machine~\cite{WaybackMachine}, a tool used for discovering internal pages within a domain, and search the keyword \textit{``[company name/model name] + privacy policy/privacy notice''} in Google search to identify any documents missed during the manual browsing stage. We further collect different versions of documents caused by jurisdiction, such as OpenAI's Europe privacy policy~\cite{OpenAIEUPP}.
After constructing an initial list of documents, we include documents that are related to privacy.
% For example, privacy FAQs and technology explanation articles.
Notably, we exclude the following categories of documents:
\begin{itemize} [leftmargin=*, noitemsep, topsep=3pt]
\item Cookie policies
% \item Security policies
\item Usage policies
\item Privacy centers or landing pages
\item Privacy policy summaries
\end{itemize}

% As we focus on the privacy policy of LLMs, we do not collect the general privacy policies of the whole group like Google and Meta because the scope they describe is the large business of the whole company, not specifically AI services. 
We find the scope of privacy policies for Google and Meta is the large business of the whole company, not specifically their AI services. Since our analysis focus on privacy policies for LLMs, we consider that such holistic privacy policies for general services might introduce too much non-AI-related content to contaminate the analysis. We therefore do not include them in our dataset.
% \zhen{@Shidong please check.}
% \sd{Having those privacy polices might contaminate the analysis.} 
Instead, we collect their supplemental documents for AI services, such as Gemini Apps Privacy Hub~\cite{GeminiAppsPrivacyHub} and Meta AIs Terms of Service~\cite{MetaAIsTerms}.
In total, our document list consists of 10 privacy policies and 52 privacy-related supplemental documents available at:
~\url{https://github.com/ZhenTAO3059/A-Longitudinal-Measurement-of-Evolution-in-Privacy-Policies-for-Large-Language-Models}.
% ~\url{https://anonymous.4open.science/r/A-Longitudinal-Measurement-of-Evolution-in-Privacy-Policies-for-Large-Language-Models-66E7}.

% \zhen{add description and example about document theme.}
% \sd{@Chunyang Do we need to say we have a repo for those docs?}
% \sd{yes we do}

% https://anonymous.4open.science/r/A-Longitudinal-Measurement-of-Evolution-in-Privacy-Policies-for-Large-Language-Models-66E7/README.md

\subsection{Chronological Dataset}
We leverage Wayback Machine~\cite{WaybackMachine}, a popular archive of snapshots of Internet sites commonly used in longitudinal analysis~\cite{adhikari2023evolution, lovato2023more, amos2021privacy, wagner2023privacy}, to collect historical versions of the documents through the URLs. 
We observe that privacy policy URLs may change over time, we therefore use the site map of the base URL to identify the location of the privacy policy at past points in time. 
For example, through reviewing the site map of \textit{\url{https://openai.com/}}, we find the privacy policy was moved from a previous directory (\texttt{/api/policies/privacy/})\footnote{\href{https://openai.com/api/policies/privacy/}{https://openai.com/api/policies/privacy/}} 
% \footnote{\href{https://web.archive.org/web/20220323134726/https://openai.com/api/policies/privacy/}{https://web.archive.org/web/20220323134726/https://openai.com/api/policies/privacy/}} 
% \footnote{%
%   \href{https://web.archive.org/web/20220323134726/https://openai.com/api/policies/privacy/}%
%        {\nolinkurl{https://web.archive.org/web/20220323134726/https://openai.com/api/policies/privacy/}}%
% }
to the current directory (\texttt{/policies/privacy-policy/}).\footnote{\href{https://openai.com/policies/privacy-policy/}{https://openai.com/policies/privacy-policy/}} For each document, we utilize Waybackpack~\cite{Waybackpack} to automatically crawl the archive and download the snapshots captured by Wayback Machine, with their capture timestamps in the form of ``yyyyMMddHHmmss''. To ensure reliability, each time an error occurs while crawling the HTML content of a web page, such as a 4XX or 5XX status code, a maximum of three retries are performed, with an interval of 5 seconds between each retry. Considering the capture by Wayback Machine and the crawling by Waybackpack are not always successful, we perform a validity check on each downloaded file and remove failed crawls. 
% Specifically, we use the Beautiful Soup  Python library~\cite{beautifulsoup} to extract the HTML content and filter out documents with empty content or less than xx \chen{??} words. \chen{For the implementation, it is not necessary to mention such details like introducing what libarary do you use?}

% There exists redundant captures with the same content. To identify the changes in privacy policies over time, we further compare the contents of two consecutive captures, and screen out a sub-dataset only. 
We then remove the duplicate captures and only keep the first of each document with its time stamp.
Specifically, we extract the content-only text from each document by parsing the page, removing boilerplate elements (e.g., $<$footer$>$) to minimize noise.
We implement a document change extraction mechanism leveraging sentence similarity to identify the textual differences between each consecutive policy version. We further manually annotate each identified change into our taxonomy categories to understand what content is evolving. We present more details in Section~\ref{PP_analysis_method}. 
% In addition, we also record the update time displayed in the document, e.g., ``last update date''. We observed that the document versions indicated by changes in the ``last update time'' do not completely match to the changes we extracted based on text. We discuss this further in Section~\ref{rq3_update_frequency}. 
With this pipeline, we collect unique versions of archived privacy-related documents with 3,463 annotated policy edits. 
% With this pipeline, we collect archived privacy-related documents of 11 mainstream LLM providers, including 11 privacy policies and 51 privacy-related additional documents. 

% \input{Tables/LLM_providers}

 %-----------------
\begin{figure*}[t]
  \centering
  \includegraphics[width=.98\linewidth]{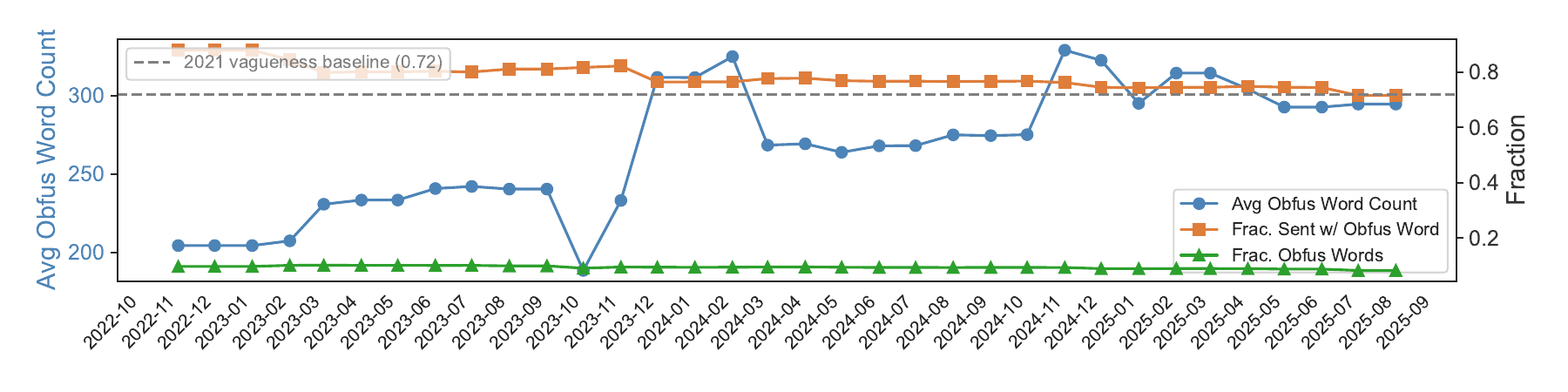}
  \caption{Obfuscating word count, fraction of obfuscating word, and fraction of sentences with obfuscating words in policies.}
  \label{fig_vagueness}
\end{figure*}
%-----------------

\subsection{Privacy Policy Analysis}
\label{PP_analysis_method}
Privacy policies are increasingly lengthy, and the readability and vagueness of these long documents has been a long-standing question~\cite{amos2021privacy, shipp2020private, wagner2023privacy, adhikari2023evolution, liu2016modeling, singh2011evaluating}. 
We therefore conduct a comprehensive evaluation on those aspects. 
To observe how policies evolve and understand the influencing factors, we extract changes between versions and collect key events related to LLM providers.

\textbf{Readability and Vagueness Calculation.} Following the approach in prior work on privacy policy analysis~\cite{amos2021privacy}, we examine readability with the Flesch-Kincaid grade level (FKGL)~\cite{kincaid1975derivation},
% through the Python library \textit{py-readability-metrics}~\cite{readabilitymetrics} \chen{too detailed.}, 
which represents grade level in the U.S. school system and is widely used in privacy policy analysis~\cite{adhikari2023evolution, xie2025evaluating, amos2021privacy}.
Measuring obfuscating words is commonly used in language accessibility~\cite{shipp2020private, wagner2023privacy}. Following previous work~\cite{wagner2023privacy}, we leverage the list of obfuscating adjectives and adverbs published by~\cite{shipp2020private} to measure the vagueness of the policy text. We apply gensim preprocessor~\cite{Gensim} and PorterStemmer algorithm~\cite{porter1980algorithm} to preprocess the policy text and the list of obfuscating words, such as normalizing text and stripping punctuations. The vagueness sentence fraction ranges from 0 to 1.

\textbf{Policy Change Extraction.} In the longitudinal analysis of policies, we aim to discover patterns from policy content evolution over time. To extract sentence-level edits between consecutive policy versions, we apply a lightweight heuristic. 
First, each version is normalized and split into sentences. We then align the old and new sentences: (1) exact matches are ignored; (2) each unmatched old sentence is paired with the most similar unused new sentence using Python library \textit{difflib.SequenceMatcher}~\cite{difflib} and labeled \textbf{[modify]} if the similarity score is higher than 0.5, otherwise \textbf{[delete]}; (3) any remaining new sentences are labeled \textbf{[add]}. 
% Our choice of 0.5 is close to the default ``good enough'' cutoff of 0.6 in \textit{difflib} used for near-duplicate detection \chen{\cite{??}}. We lower it slightly to avoid missing edits and then validate with human review. 
We conduct a pilot study to determine the best similarity threshold after manually examining the results.
% \chen{Too engineering for this paragraph. You can directly say that you carry out some pilot study to dermine the best parameter after the manual check.} \sd{agree} 
Finally, two of the authors manually annotate each detected edit into our taxonomy categories and run a sanity check of the annotations. Annotators reach 92.8\% raw agreement (Cohen’s $\kappa = 0.92$). Most of policy edits belong to \textbf{[add]} category and often appear in groups, which reduces the difficulty of annotation to some extent and leads to a high Cohen’s $\kappa$.
% \zhen{@Shidong please check.} \sd{explain the difficulkty of the annotation task} 
All cases of disagreement are subsequently discussed to reach a final consensus annotation.
% \sd{inter-rater agreement}

\textbf{Construction of the LLM Key Event Timeline.} After capturing the content evolution, to understand the motivations behind these changes, we collect key events related to LLMs in the recent three years. Following previous work~\cite{ali2025understanding}, we manually review the events from Google search results. We also enable ChatGPT's web search feature to look for news blogs in the LLM ecosystem. In addition, as GDPR is a benchmark regulation for protecting data privacy, we search for legal cases involving LLM companies in the GDPRhub~\cite{GDPRhub}, a commonly used database collecting GDPR-related legal decisions. After verifying events with the articles on the official websites of LLM providers, we result in 18 key events, including product releases (e.g., model launches) and regulatory actions (e.g., investigations from data protection authorities).
% , and corporate development (e.g., establishment of partnerships). 
We do not aim to be exhaustive of all events or to force policy changes to key events, but aim to briefly understand how LLM privacy policies may be impacted by observing the timeline.

\section{Differences between Privacy Policies of LLMs and General Software (RQ1)}
\label{sec_rq1}
In this section, %we investigate the differences between privacy policies for LLMs and traditional software. 
we analyze the differences between privacy policies for LLMs and those for general software systems. 
% \chen{This kind of sentence appear several times in the paper, so please pay attention to its grammar. ``between policy for LLm and \textbf{those for}'', ``traditional'' to ``general''. If you agree, please update other parts in the paper.}
General software refers to mobile apps, websites, and other software that are not LLM-based applications. 
% We present a stylometric analysis regarding the length, readability and vagueness of policy text and compare the results with previous studies on privacy policies for general software. We summarize the unique patterns and propose a new taxonomy tailored to LLM privacy policies, highlighting the LLM-specific policy contents. 

\subsection{Stylometric Analysis}

% \textbf{Policy Length.} 
% LLM providers’ policies are substantially long. As shown in Table~\ref{tab_LLM_provider}, every LLM provider's average policy length exceeds 2,000 words, with a mean of 3,346 words across all versions. Notably, xAI's policies have an average length of 5,646 words. In comparison, the average length of traditional software policies is 2,178 words in March 2019~\cite{wagner2023privacy, amos2021privacy}. The median LLM policy length is 2,901 words, significantly exceeding the traditional software's policies (1,522 words in 2019)~\cite{amos2021privacy}. Considering software popularity, higher-ranked policies are reported to be longer than lower-ranked policies~\cite{libert2018automated}, with an average length of 3,200 words in 2018~\cite{wagner2023privacy}. LLM privacy policies are slightly longer than that of high-ranked software.\zhen{use xx\% to compare.}
% \chen{suggest to introduce the study for the comparison here.}
Length, readability and vagueness of privacy policies are frequently measured in researches~\cite{wagner2023privacy, amos2021privacy, adhikari2023evolution, xie2025evaluating, shipp2020private}. We therefore present a stylometric analysis regarding these aspects and compare the results with previous studies on privacy policies for general software, using the statistics prior to the LLM era as a historical baseline rather than a contemporaneous control. Our goal is to provide an intuitive yardstick for the magnitude of LLM policy features, since post-2022 policies might embed AI features and would not constitute a general software comparison group. 
% To reduce bias, we report both absolute statistics and effect sizes relative to the baseline.\zhen{Please check whether this claim is fine.}

\textbf{Policy Length.} 
LLM providers’ policies are substantially long. As shown in Table~\ref{tab_LLM_provider}, every LLM provider's average policy length exceeds 2,000 words, with a mean of 3,346 words across all versions, which is around 53.6\% longer than the average length of general software policies in March 2019 (2,178 words)~\cite{wagner2023privacy,amos2021privacy}. Notably, xAI’s policies reach 5,646 words in average. The median LLM policy length is 2,901 words and is about 90.6\% longer than the reported 2019 median for general software (1,522 words)~\cite{amos2021privacy}. 
% \chen{One potential attack point of using this work is that this paper is too old, and it may not be fair to compare 2022-2025 LLM and <2019 general software, as the privacy policy for general software in 2022-2025 may be different from those before 2019.} 
Considering software popularity, higher-ranked software has longer policies than lower-ranked ones~\cite{libert2018automated}, reporting 3,200 words on average in 2018~\cite{wagner2023privacy}.
%LLM privacy policies are even slightly longer than those of high-ranked software. 
%Even relative to higher-ranked software baseline, LLM policies are still slightly longer on average. 
Overall, LLM privacy policies are longer than general software policies, including those of higher-ranked software.

% Interestingly, we observed that the policy length does not always increase gradually, but sometimes there is a significant change in the word count, as shown in the two cutoff examples in Figure~\ref{fig_word_count_bar}...We discuss such evolution further in~\ref{}...

% %-----------------
% \begin{figure}[t]
%   \centering
%   \includegraphics[width=.99\linewidth]{Figures/new_word_count_bar_comparison.pdf}
%   \caption{Word count of privacy policies for LLMs in January 2024 (green) and November 2024 (blue). The horizontal line represents the 2019 baseline word count (1,522)~\cite{amos2021privacy}.}
%   \label{fig_word_count_bar}
% \end{figure}
% %-----------------

\textbf{Readability.}
As shown in Table~\ref{table_rv_statistics}. The median FKGL rises from 14.77 (2023) to 15.66 (2024) and 15.74 (2025), which are well above the historical medians for tradition software policies (11.9 in 2000 and 13.2 in 2019)~\cite{amos2021privacy}. A median over 14 implies the LLM privacy policies require upper-college to graduate-level reading ability to understand the contents. 
% We present the distribution of FKGL median over time in Figure~\ref{fig_readability}. The entire distribution sits above 14 over the recent 3 years, confirming that high reading difficulty remains persistent. In 2022, there were fewer LLM providers offering services, e.g., OpenAI, and the scores therefore reflect the high reading difficulty of these early privacy policies. 
% A soft dip toward 14 is seen in early 2023 as more providers joined, and there is a re-acceleration through 2024-2025 with medians around 15 and greater dispersion. This widening spread suggests uneven editorial practices across revisions, which we discuss further in Section~\ref{RQ3_content_evo}.
Overall, LLM policies remain consistently hard to read and are trending slightly more complex.

% %-----------------
% \begin{figure*}[t]
%   \centering
%   \includegraphics[width=.98\linewidth]{Figures/new_readability_boxplot.pdf}
%   \caption{FKGL Scores of privacy policies for LLMs. \chen{Do not see the big differece, and table 2 overlap a bit with this. So may remove this to save some space?}}
%   \label{fig_readability}
% \end{figure*}
% %-----------------

\begin{table}
\scriptsize
\caption{The median FKGL score and average vagueness sentence fraction by year. Data for 2025 is as of August. 
% \sd{as one is absolute value and another is a ratio value (0-1). so we might need to add a +/- \% to indicate how to interpret the delta.}
% \chen{Why should we have $\downarrow$?}
}
\label{table_rv_statistics}
% \begin{minipage}{\textwidth}
% \begin{subtable}{0.5\textwidth}
% \centering
% \begin{tabular}{lccc}
% \toprule
% \textbf{} & \textbf{gpt3demo} & \textbf{gpt4demo} & \textbf{Total} \\ \midrule
% Available Code Repository & 138 & 10 & 148 \\
% %Relevant Website & 60 & 7 & 67 \\
% Available Privacy Policy & 16 & 2 & 18 \\ 
% Available Privacy Labels & 1 & 1 & 2 \\ 
% \midrule
% GAI Tools & 887 & 87 & 974 \\ 
% \bottomrule
% \end{tabular}
% \subcaption{Overall.}
% \end{subtable}
% \end{minipage}
\begin{minipage}{\textwidth}
\begin{subtable}[t]{0.25\textwidth}
\centering
\scriptsize
\label{table_read_statistics}
\begin{tabular}{lc}
\toprule
\textbf{Year} & \textbf{Median FKGL Score} \\ 
\midrule
2025 & 15.74 \\ 
2024 & 15.66 \\
2023 & 14.77 \\
\midrule
2019 &\textbf{13.20}~\cite{amos2021privacy} \\
\bottomrule
\end{tabular}
% \subcaption{Privacy policies.}
\end{subtable}
% \hspace{-0.02\textwidth}
\begin{subtable}[t]{0.25\textwidth}
\centering
\label{table_vague_statistics}
\begin{tabular}{lc}
\toprule
\textbf{Year} & \textbf{Average Vagueness} \\ 
\midrule
2025 & 0.75 \\ 
2024 & 0.76 \\
2023 & 0.81 \\
\midrule
2021 & \textbf{0.72}~\cite{wagner2023privacy} \\
\bottomrule
\end{tabular}
% \subcaption{GitHub repositories.}
\end{subtable}
\end{minipage}
\end{table}

% Median FKGL Score by Year (2023-2025):
%    year  median_fkgl_score
% 0  2023          14.883615
% 1  2024          15.902360
% 2  2025          16.114088

% Average Vagueness Fraction (obfus_sent_frac) by Year (2023-2025):
%    year  avg_vagueness_frac
% 0  2023            0.806443
% 1  2024            0.767112
% 2  2025            0.759665

\textbf{Vagueness.}
%  72\% of policy sentences in 2021 contained at least one obfuscating word
% ``Certain actions you take may be visible to other users of the Service'' (added in OpenAI 20230227230602 and deleted in 20230314210410) in \textit{Sharing and disclosure of personal information} Section.
The average fraction of sentences containing at least one obfuscating word is 0.72 in 2021 for general software policies~\cite{wagner2023privacy}, while LLM policies' value peaks at 0.81 in 2023, then declines modestly to 0.76 (2024) and 0.75 (2025). As presented in Figure~\ref{fig_vagueness}, this fraction starts high in 2022, with a peak of 0.81 in 2023, and gradually falls toward 0.75 afterwords. That is, vagueness of policy text remains widespread, with roughly 75\% of sentences still containing at least one obfuscating word, such as ``\textit{might}'' or ``\textit{sometimes}''. The average obfuscating word count rises from around 200 to around 300 in 2025, indicating a growing absolute volume of indeterminate language. The fraction of obfuscating words stays remarkably flat, since the absolute policy length also increases slightly over time. 
% \chen{May give some examples of these words, especially LLM specific?}

LLM-specific disclosures drives challenges on informing. Policies may include technical nouns and additional descriptions in an attempt to reduce vagueness, but at the expense of lower readability. 
% One sentence is shown in Figure~\ref{fig_policy_excerpt}. 
Moreover, we observe many LLM policies link out to supplementary documents, such as notices about model training. The reading burden then increases when users must chase fragmented information across multiple documents, raising the practical cost of informed consent. Overall, LLM privacy policies are beyond college-level readability and is more vague than general software policies, even as vagueness falls slightly after 2023. 
We believe LLM providers are likely aware of these problems, as Anthropic~\cite{anthropicdocs} and OpenAI~\cite{OpenAIPP} have introduced the LLM-empowered inquiry feature for sprawling documents that can help users search or understand the contents, aiming to reduce the cost of comprehension. 

Several factors may have driven this vagueness. First, the black-box characteristics of AI, such as non-deterministic behaviors, makes precise statements about data difficult without oversimplifying. Second, regulatory uncertainty and heterogeneity surrounding AI may push providers toward de-specified commitments to keep one policy viable across jurisdictions, which we discuss further in Section~\ref{rq2_Juris_sec}. Third, rapid product iteration for LLM services may incentivize flexible language over concrete statements that could be invalid within iterations. Our findings in Section~\ref{rq3_driver_sec} also show that product release is a policy-change driver.

\subsection{Characterization Observation}~\label{rq1_patterns}
We present observations on LLM privacy policies that distinguish them from policies for general software. These shifts in content reflect changes in privacy practices in the context of LLM era and also reveal new privacy challenges. 
% \chen{Why do you analyze the difference from these aspects? Any criteria? In addition, I do not see the comparison with general software in the first few aspects, but more about LLM privacy policy. So should we put this section as one RQ?}

\textbf{First-party Data Collection.} 
LLM privacy policies broaden first-party collection to explicitly include chat transcripts, AI-generated outputs, and user feedback on model responses as personal data. Providers describe collection from a collector’s perspective, being transparent that they collect data not just from users but from broader sources enumerated in three groups: users-provided, automatically collected, and received from other sources, which often includes publicly available internet sources and third parties such as Github~\cite{Github} and Hugging Face~\cite{HuggingFace}. They often acknowledge that publicly available data on the internet may contain personal data, e.g., ``\textit{we understand that there is personal information incidentally included in these datasets}'' (P3).\footnote{In the rest of the paper, we alternatively use (P1) to (P11) to indicate the LLM provider as enumerated in the Table~\ref{tab_LLM_provider} for the simplicity.} Interestingly, possibly due to the difficulty of filtering, providers do not seem to be confident in their actions. P4 once claimed they \textit{``use \underline{best practices} to filter out''}, but downgraded to \textit{``\underline{good practices}''} in the later version. Overall, disclosures are increasingly source-specific and modality-specific. Policies also introduce model training and research as collection purposes, reflecting providers’ scientific research orientation, while downplaying advertising, as current LLM services do not rely on targeted ads for monetization~\cite{vekaria2025big}). However, Gemini, ChatGPT, and Perplexity are gradually exploring advertising features~\cite{GoogleAds, OpenAIAds, perplexityAds}.

\textbf{Personal Information Reproduction and Inferring.}
LLM providers explicitly warn of a novel privacy risk that models may remember and reproduce personal information users previously supplied in the input, blurring the line between provided and derived data. Studies show LLMs can overfit to training data and inadvertently expose sensitive content~\cite{gupta2023chatgpt, kibriya2024privacy, li2023privacy}, especially under adversarial prompts~\cite{yu2023gptfuzzer}, and can infer personal attributes with high accuracy (85\%)~\cite{staab2023beyond}. Aware of this, several providers assert they do not infer user characteristics, e.g., ``xAI does not process training data for the purposes of \underline{inferring or deriving} any sensitive or special category data about individuals'' and ``we do not process sensitive Personal Data for the purposes of \underline{inferring characteristics}''(P3). However, we observe wavering language over time. For instance, P1 once claimed \textit{``we may analyze the general behavior and characteristics of users of our Services...''}, but removed the relevant clauses in later policy updates. These disclosures underline the persistence of generative context. Information like a user’s name can resurface long after input, challenging user privacy expectations combined with limited user awareness of LLM memory mechanisms~\cite{zhang2024ghost}. A transient data interaction in general software can become a retained and resurfaced piece of data in the LLM context.

\begin{table*}[t]
\caption{Taxonomy of privacy policies for LLMs. The \emph{attributes} under each \textbf{category} are \emph{in italics}, and new attributes in addition to the OPP-115 taxonomy~\cite{wilson2016creation} are marked with \newstar.
% \chen{what does it mean? New added compared with privacy policy for general software? If so, make it clear.} 
The \capbg{grpRC}{blue attributes} indicate that they have new values in the LLM scenario, which is discussed in Section~\ref{rq1_patterns}. The taxonomy is partially presented, and the complete version can be found in our dataset.
% The taxonomy is partially presented, and the complete taxonomy can be found in the Appendix Section~\ref{append_taxonomy}. 
% \chen{Can we make it to take the whole page by including all info, no need to put it to Appendix.}
% \chen{1. missing some prevalence data for some attributes, zero? 2. Can we make it to take the whole page by including all info, no need to put it to Appendix. 3. For attribute, can you nest it a bit to make it look more like kids of each attribute name?}
% New categories are labeled with \newstar, including \capbg{grpRC}{LLM-specific} and \capbg{grpMD}{general modifications}.
}
\label{tab_taxonomy_short}
\centering
\resizebox{.98\textwidth}{!}{%
\begin{tabular}{
    % >{\raggedright\arraybackslash}p{3.0cm}  % High-level
    % >{\raggedright\arraybackslash}p{5.0cm}  % Second-level
    % >{\raggedright\arraybackslash}p{5.0cm}  % Explanation
    % >{\centering\arraybackslash}p{3.0cm}    % Prevalence
    l|l|l|ccccc|ccccc
}
\toprule
 % & \textbf{Category} & \textbf{Explanation} & \textbf{Prevalence (\%) of LLM Providers} \\
 & \textbf{Category} & \textbf{Explanation} & \textbf{P1} & \textbf{P1(EU)} & \textbf{P2} & \textbf{P3} & \textbf{P4} & \textbf{P5} & \textbf{P6} & \textbf{P7} & \textbf{P8} & \textbf{P9}\\
\midrule
% \rowcolor[]{}
\multirow{24}{*}{\textbf{Data Practices}}
  & \textbf{First-party Data Collection} & Data collection or data use by the company/organization owning the service. & \multicolumn{10}{c}{\cellcolor{lightgray!35}{}} \\
  \cmidrule(lr){2-13}
  & -- \emph{Collection Mode}                   & If the data collection is implicit or explicit. & \CIRCLE &\CIRCLE &\CIRCLE &\CIRCLE &\CIRCLE &\CIRCLE &\CIRCLE &\CIRCLE &\CIRCLE &\CIRCLE \\
  \cmidrule(lr){2-13}
  & \cellcolor{grpRC}-- \emph{Action First-Party}                   & How the first party collects, tracks, or obtains user information. & \CIRCLE &\CIRCLE &\CIRCLE &\CIRCLE &\CIRCLE &\CIRCLE &\CIRCLE &\CIRCLE &\CIRCLE &\CIRCLE \\
  \cmidrule(lr){2-13}
  & -- \emph{Identifiability}                   & Whether data is aggregated, anonymized, or de-identified before being used. & \CIRCLE &\CIRCLE &\CIRCLE &\CIRCLE &\CIRCLE &\Circle &\CIRCLE &\CIRCLE &\Circle &\CIRCLE \\
  \cmidrule(lr){2-13}
  & \cellcolor{grpRC}-- \emph{Information Type}                   & What category of information is collected or tracked by the company/organization. & \CIRCLE &\CIRCLE &\CIRCLE &\CIRCLE &\CIRCLE &\CIRCLE &\CIRCLE &\CIRCLE &\CIRCLE &\CIRCLE \\
  \cmidrule(lr){2-13}
  & \cellcolor{grpRC}-- \emph{Purpose}                   & The purpose of collecting or using user information. & \CIRCLE &\CIRCLE &\CIRCLE &\CIRCLE &\CIRCLE &\CIRCLE &\CIRCLE &\CIRCLE &\CIRCLE &\CIRCLE \\
  \cmidrule(lr){2-13}
  & -- \emph{User Type}                   & If this practice applies specifically to users w/o an account. & \Circle &\Circle &\CIRCLE &\Circle &\CIRCLE &\Circle &\Circle &\Circle &\Circle &\Circle \\
  \cmidrule(lr){2-13}
  & \textbf{Third Party Sharing/Collection}             & Data sharing with third parties or data collection by third parties. & \multicolumn{10}{c}{\cellcolor{lightgray!35}{}} \\
  \cmidrule(lr){2-13}
  & \cellcolor{grpRC}-- \emph{Scenario}\newstar                   & The circumstances of interaction with third parties. & \CIRCLE &\CIRCLE &\CIRCLE &\CIRCLE &\CIRCLE &\CIRCLE &\CIRCLE &\CIRCLE &\CIRCLE &\CIRCLE \\
  \cmidrule(lr){2-13}
  & \cellcolor{grpRC}-- \emph{Third Party Entity}                   & The third-party involved in the data practice. & \Circle &\Circle &\Circle &\CIRCLE &\Circle &\CIRCLE &\Circle &\Circle &\CIRCLE &\CIRCLE \\
  \cmidrule(lr){2-13}
  & \cellcolor{grpRC}-- \emph{Action Third Party}                   & How the third-party receives, collects, tracks, or sees user information. & \CIRCLE &\CIRCLE &\CIRCLE &\CIRCLE &\CIRCLE &\CIRCLE &\CIRCLE &\CIRCLE &\CIRCLE &\CIRCLE \\
  \cmidrule(lr){2-13}
  & \cellcolor{grpRC}-- \emph{Information Type}                   & What category of information is collected or tracked by the third-party. & \Circle &\Circle &\Circle &\CIRCLE &\Circle &\Circle &\Circle &\CIRCLE &\Circle &\CIRCLE \\
  \cmidrule(lr){2-13}
  & \cellcolor{grpRC}-- \emph{Purpose}                   & The purpose of a third party receiving or collecting user information. & \CIRCLE &\CIRCLE &\CIRCLE &\CIRCLE &\CIRCLE &\CIRCLE &\CIRCLE &\CIRCLE &\CIRCLE &\CIRCLE \\
  \cmidrule(lr){2-13}
  &-- \emph{User Type}                   & If this practice applies specifically to users w/o an account. & \Circle &\Circle &\CIRCLE &\Circle &\Circle &\Circle &\Circle &\Circle &\Circle &\Circle \\
  \cmidrule(lr){2-13}
  & \cellcolor{grpRC}-- \emph{Policy Application Scope}\newstar                   & Whether and to what extent the privacy policy continues to apply. & \CIRCLE &\CIRCLE &\CIRCLE &\CIRCLE &\CIRCLE &\CIRCLE &\CIRCLE &\CIRCLE &\CIRCLE &\CIRCLE \\
\midrule
\multirow{12}{*}{\textbf{\makecell[l]{User Rights\\ \& Controls}}}
  & \textbf{User Choice/Control}             & General choices and control options available to users. & \multicolumn{10}{c}{\cellcolor{lightgray!35}{}}  \\
  \cmidrule(lr){2-13}
  & \cellcolor{grpRC}-- \emph{Choice Scope}                   & What scope does the user choice or control apply to. & \CIRCLE &\CIRCLE &\CIRCLE &\CIRCLE &\CIRCLE &\CIRCLE &\CIRCLE &\CIRCLE &\Circle &\CIRCLE \\
  \cmidrule(lr){2-13}
  & \cellcolor{grpRC}-- \emph{Information Type}                   & What category of information does the user choice apply to. & \Circle &\Circle &\Circle &\Circle &\Circle &\Circle &\Circle &\Circle &\Circle &\CIRCLE \\
  \cmidrule(lr){2-13}
  & \cellcolor{grpRC}-- \emph{Purpose}                   & What purpose/use of information does the user choice apply to. & \CIRCLE &\CIRCLE &\Circle &\CIRCLE &\CIRCLE &\CIRCLE &\CIRCLE &\Circle &\CIRCLE &\CIRCLE \\
  \cmidrule(lr){2-13}
  &-- \emph{User Type}                   & If this practice applies specifically to users w/o an account. & \Circle &\Circle &\Circle &\Circle &\Circle &\Circle &\Circle &\Circle &\Circle &\Circle \\
  \cmidrule(lr){2-13}
  & \textbf{User Access, Edit and Deletion} & Users' rights to access, edit or delete the data that the company has about them.  & \multicolumn{10}{c}{\cellcolor{lightgray!35}{}} \\
  \cmidrule(lr){2-13}
  & \cellcolor{grpRC}-- \emph{Access Type}                   & Options offered for users to access, edit, delete user information. & \CIRCLE &\CIRCLE &\CIRCLE &\CIRCLE &\CIRCLE &\CIRCLE &\CIRCLE &\CIRCLE &\CIRCLE &\CIRCLE \\
  \cmidrule(lr){2-13}
  &-- \emph{User Type}                   & If this practice applies specifically to users w/o an account. & \Circle &\Circle &\CIRCLE &\Circle &\CIRCLE &\Circle &\Circle &\Circle &\Circle &\Circle \\
\midrule
\multirow{9}{*}{ \textbf{\makecell[l]{Model\newstar}}}
  & \cellcolor{grpRC}\textbf{Model Training}\newstar                  & How user data is used to train or fine-tune AI models. & \multicolumn{10}{c}{\cellcolor{lightgray!35}{}}  \\
  \cmidrule(lr){2-13}
  & \cellcolor{grpRC}-- \emph{Data Sources}\newstar                   & The origins of data used to train or fine-tune AI models. & \Circle &\Circle &\CIRCLE &\Circle &\CIRCLE &\CIRCLE &\Circle &\CIRCLE &\Circle &\CIRCLE \\
  \cmidrule(lr){2-13}
  & \cellcolor{grpRC}-- \emph{Information Type}\newstar                   & What type of data is used to train AI models. & \Circle &\Circle &\CIRCLE &\Circle &\CIRCLE &\CIRCLE &\Circle &\CIRCLE &\Circle &\CIRCLE \\
  \cmidrule(lr){2-13}
  & \cellcolor{grpRC}-- \emph{Identifiability}\newstar                   & Whether data is aggregated, anonymised, or de-identified before being used. & \CIRCLE &\CIRCLE &\CIRCLE &\CIRCLE &\Circle &\CIRCLE &\Circle &\CIRCLE &\Circle &\CIRCLE \\
  \cmidrule(lr){2-13}
  & \cellcolor{grpRC}-- \emph{Opt-out for Model Training}\newstar                   & If user can refuse or disable the use of their data for model training. & \CIRCLE &\CIRCLE &\Circle &\Circle &\CIRCLE &\Circle &\Circle &\Circle &\Circle &\Circle \\
  \cmidrule(lr){2-13}
  & \cellcolor{grpRC}-- \emph{Supplementary Notice}\newstar                   & Whether supplemental document are provided to explain the training detail. & \CIRCLE &\CIRCLE &\CIRCLE &\CIRCLE &\CIRCLE &\CIRCLE &\Circle &\Circle &\CIRCLE &\Circle \\
\midrule
\multirow{3}{*}{\textbf{\makecell[l]{Governance\\ \& Compliance}}}
  & \textbf{Policy Change}   & If and how users will be informed of policy changes. & \multicolumn{10}{c}{\cellcolor{lightgray!35}{}} \\
  \cmidrule(lr){2-13}
  & \textbf{International and Specific Audiences}        & Specific audiences that may be offered special provisions. & \multicolumn{10}{c}{\cellcolor{lightgray!35}{}}  \\
\midrule
\multirow{10}{*}{\textbf{\makecell[l]{Data\\ \& Management}}}
  & \textbf{Data Retention} & The retention period for collected user information. & \multicolumn{10}{c}{\cellcolor{lightgray!35}{}}  \\
  \cmidrule(lr){2-13}
  &-- \emph{Retention Period}                   & Description of the retention period. & \CIRCLE &\CIRCLE &\CIRCLE &\CIRCLE &\CIRCLE &\CIRCLE &\Circle &\CIRCLE &\Circle &\CIRCLE \\
  \cmidrule(lr){2-13}
  & \cellcolor{grpRC}-- \emph{Retention Purpose}                   & The purpose to which the retention practice applies. & \CIRCLE &\CIRCLE &\CIRCLE &\CIRCLE &\CIRCLE &\CIRCLE &\Circle &\CIRCLE &\CIRCLE &\CIRCLE \\
  \cmidrule(lr){2-13}
  & \cellcolor{grpRC}-- \emph{Information Type}                   & The information type for which the retention period is specified. & \Circle &\Circle &\Circle &\Circle &\CIRCLE &\Circle &\Circle &\Circle &\Circle &\Circle \\
  \cmidrule(lr){2-13}
  & \textbf{Data Security}                   & How user information is secured and protected.  & \multicolumn{10}{c}{\cellcolor{lightgray!35}{}} \\
  \cmidrule(lr){2-13}
  &-- \emph{Data Transfer}                   & Cross-border or other onward transfers of personal data. & \CIRCLE &\CIRCLE &\CIRCLE &\Circle &\CIRCLE &\CIRCLE &\CIRCLE &\CIRCLE &\CIRCLE &\CIRCLE \\
  \cmidrule(lr){2-13}
  &-- \emph{Data Storage}                   & Where, and under whose control, user data is stored. & \CIRCLE &\CIRCLE &\CIRCLE &\CIRCLE &\CIRCLE &\CIRCLE &\Circle &\CIRCLE &\Circle &\CIRCLE \\
\bottomrule
\end{tabular}
}
\end{table*}

\textbf{Third-party Data Interactions.} 
LLM privacy policies flag third-party data interactions that differ from those of general software, adding two notable categories: (a) plugins/integration partners and (b) downstream applications utilizing LLM API services. Policies often explicitly cover these extension scenarios, explaining that user queries and related data may be sent to external services to fulfill requests, e.g., ``\textit{you can also send information to third-party applications, such as via \underline{custom actions for GPTs}, or for \underline{searching the web}...}'' (P1). Providers set their policy application scope for third-party scenarios. For instance, P5 states its policy does not cover data collection when accessing ``\textit{\underline{downstream systems or applications} developed by developers using our open platform services},” and “\textit{this Privacy Policy does not apply where Anthropic acts as a data processor},'' such as ``\textit{you're using an app that is powered on the back-end with Claude}'' (P2). It is reported that such downstream GenAI assistants may also infer user attributes~\cite{vekaria2025big}. Notably, human annotators that read and annotate the user conversations are also treated as third parties in Gemini Apps Privacy Notice, Google's supplementary document for Gemini, which warns users not to include information they ``\textit{wouldn’t want a reviewer to see}'' (P11). P4 further treats ``\textit{links, thumbnails, or snippets}'' as ``\textit{Third-Party Content}'', noting that “\textit{if you include personal data in your Input, personal data might also appear in the Third-Party Content}” (P4). We regard these disclosures as a substantive shift from policies for general software, where third parties often mean background analytics and ads toward user-initiated data exchanges. 

\textbf{Use of Data for Model Training.} 
A key feature of LLM services is using user data to improve the models. Unlike general software that primarily uses data to provide or personalize a service, many LLM policies explicitly list training or improving the underlying models as a processing purpose. Policies typically justify this by aggregating or anonymizing data and stressing that training is not on an identifiable individual basis. We also observe evolution over time. After Italy’s temporary ban for alleged GDPR violations~\cite{gdprhubItaly}, P1 added more transparent disclosures and user controls for training data use in policy updates. We further discuss such evolution in Section~\ref{RQ3_content_evo}. As features expand and third-party services integrate, some providers such as P4 emphasize that user input and output accessed by third parties will not be used to train third-party AI models. User rights also differentiate by user type. P4 states that free-tier inputs and outputs are used for training, while paid-tier data is not. P2 renamed its supplemental document from ``Notice on Model Training'' to ``Non-user Privacy Policy'' to highlight the distinction between ``\textit{\underline{users}}'' and ``\textit{\underline{non-users}}''. 

\textbf{User Rights and Choices.}
LLM privacy policies reaffirm traditional rights, such as access and deletion, while extending these rights with LLM-specific choices, notably opt-outs from using data to improve models. However, these rights often include exceptions. P2 states that if a conversation is flagged by its safety team, it will be used for model training. We observe providers may address response accuracy in rights sections. P1 further introduces the technical detail to users, e.g.,  \textit{``services like ChatGPT generate responses by reading a user’s request and, in response, predicting the words most likely to appear next.''}. The complexity of LLMs can limit classic rights, as outputs may include inaccurate personal data, and providers note they will ``\textit{make \underline{reasonable efforts} to correct this information}'', yet \textit{``it may not be feasible for us to do so''} (P3). Overall, transparency and control are emerging themes as providers seek to inform users about novel AI data practices without causing confusion, with a direction towards stronger user empowerment consistent with informed consent.

% LLM privacy policies generally reaffirm the user data rights covered in traditional privacy policies, such as information access and deletion, but they extend these rights with LLM-specific choices. Specifically, users are often given more granular control over how their data is utilized in model improvement, e.g., opting out. However, these rights come with specific exceptions. P2 claims that if a conversation is flagged by its safety team, it will be used for model training. In addition, 4 LLM providers include notes regarding the factual accuracy of model responses in the user right section. P1 further introduces the technical detail to users, \textit{``Services like ChatGPT generate responses by reading a user’s request and, in response, predicting the words most likely to appear next.''}. Moreover, the complexity of LLMs poses challenges to users’ traditional data rights. For instance, model’s output may contain inaccurate personal information, and although the provider will \textit{``make reasonable efforts to correct this information''}, \textit{``it may not be feasible for us to do so''}(P3). On the whole, transparency and control are becoming key themes, with providers tend to inform users about novel AI data practices and rights in a clear way without causing confusion or alarm. We believe the trajectory is toward greater user empowerment that aligns with regulatory expectations for informed consent in the age of LLMs.

\textbf{Children’s Privacy and Age Restrictions.}
LLM providers have started adopting stricter age gates than general software. For example, P9 and P2 define children to be under 18, exceeding the COPPA~\cite{COPPA} threshold of 13. This shift aligns with a broader push to harden youth protections in AI, e.g., the EU on safeguarding minors in AI-mediated environments~\cite{AIChildren}. Meanwhile, policy choices must grapple with evolving product updates. Recent news~\cite{ChatGPTerotica} indicates P1 plans to permit mature content for age-verified adults, reinforcing the need for privacy policies to clearly separate adult and children experiences and to specify data handling.

% LLM providers have started adopting stricter age-based access restrictions compared to general software. For example, P9 and P2 both require users to be at least 18 years old, a notable increase from the COPPA~\cite{COPPA} threshold of 13 years that is commonly used in conventional policies and 5 LLM providers. This shift aligns with recent researches emphasizing greater protection for youth in AI services~\cite{}.\zhen{Adult content news.}

In summary, compared to general software privacy policies, LLM policies surface disclosures that reflect conversational and multimodal interaction scenarios. This pattern could also apply to policies of other AI-enabled services that combine classic privacy commitments and AI-specific disclosures, underscoring the need for a categorization scheme. Building on these observations, we introduce a taxonomy for systematically analyzing LLM privacy policies.

\subsection{New Taxonomy}
% \zhen{based on a privacy taxonomy for website privacy policies~\cite{wilson2016creation}, we examine the privacy policies for LLMs and construct a new taxonomy tailor to LLMs. OPP-115.}
% what, why.
% \zhen{How we label some classic cases? Specific examples such as model, children, government. See how OPP-115 labels government as Data Recipient.}
LLM privacy policies describe data practices that go beyond those of general software, such as free-form prompts and data flows through external plugins. Our findings in Section~\ref{rq1_patterns} motivate us to construct a new taxonomy that can systematically capture these LLM-specific practices while remaining comparable to established work on privacy policies. Building on OPP-115, a widely used taxonomy for website privacy policies~\cite{wilson2016creation}, we propose an LLM-tailored taxonomy shown in Table~\ref{tab_taxonomy_short} that preserves OPP-115’s core structure but extends it with new content relevant to LLM data practices. Specifically, we add ``Model Training'' as a new category, including attributes regarding relevant data practices. To reflect new features in third-party data sharing and collection, we introduce attributes ``\textit{Scenario}'' and ``\textit{Policy Application Scope}''. We also extend existing attributes with LLM-specific value sets, such as new information type listed in Table~\ref{table_data_type} in the Appendix. The new taxonomy lets us systematically code both classic practices and LLM-specific behaviors within a coherent framework, enabling comparability with prior work and expressiveness for LLMs.

% What’s new (according to the taxonomy)
% New collection mode: third-party sources (e.g., Publicly available information, datasets through commercial agreements)
% “Anthropic obtains personal data from third party sources in order to train our models”
% Role transfer, from user perspective to data collector perspective
% New information type: user content, inferred information, reproduced information
% LLM-Specific Data Practices
% Blur lines between user-provided content and system-generated outputs
% New purpose: model training, scientific research
% Not individual data, but as a crowd (aggregated or de-identified information)

% What’s new (according to the taxonomy)
% Third-party products/services data interaction scenario: 
% Plugin or integration partners
% Link to or interaction with third-party services or products (TPs in search results)
% Independent chatbots (downstream systems or applications 
% Service development and user control: 
% Memory mechanism and privacy expectations
% Users may expect transparent control over how their interactions with LLMs contribute to model improvements
% Notice (providing information) -> user control
% Readability, vagueness and information overload:
% Harder to read and understand
% Reason: model training? Tech terms?
% Additional documents about data processing and model training: “Used to read one, now have to read ten”
% Need for improved communication methods or interactive tools that simplify complex terms to ensure informed consent, e.g., Inquiry Feature for OpenAI Documents

\section{Differences between Privacy Policies across Geographic Regions and LLM Providers (RQ2)}
\label{sec_rq2}

Geographic context shapes how services present privacy notices and operationalize regional requirements. Building on prior measurement of cross-region variation in policy content~\cite{ogut2024dissecting, linden2018privacy, kumar2022large, smeets2025cross}, we examine what jurisdictions and regulations LLM providers explicitly reference and most visibly intend to comply with. We find the emerging regulation de-specification pattern, then compare taxonomy coverage across providers. Finally, we observe signs of copy–paste–modify, indicating even mainstream providers could also borrow policy clauses and share templates. 

% Geographic context has been shown to shape how services present privacy notices and implement requirements from regional regulations, with measurable cross-region differences in policy contents~\cite{ogut2024dissecting, linden2018privacy, kumar2022large, smeets2025cross}. we study how LLM providers tailor their privacy policies to different regulatory environments. Specifically, we first identify which regions and regulations are explicitly referenced in LLM privacy policies that providers most visibly intend to comply with. We then analyze taxonomy coverage across mainstream LLM providers and how this coverage varies by provider and region. \chen{Too much emphasize on region, rather than different providers and other aspects in this paragraph.}

\subsection{The Impact of Territorial Jurisdictions}
\label{rq2_Juris_sec}

%-----------------
\begin{figure*}[t]
  \centering
  \includegraphics[width=.98\linewidth]{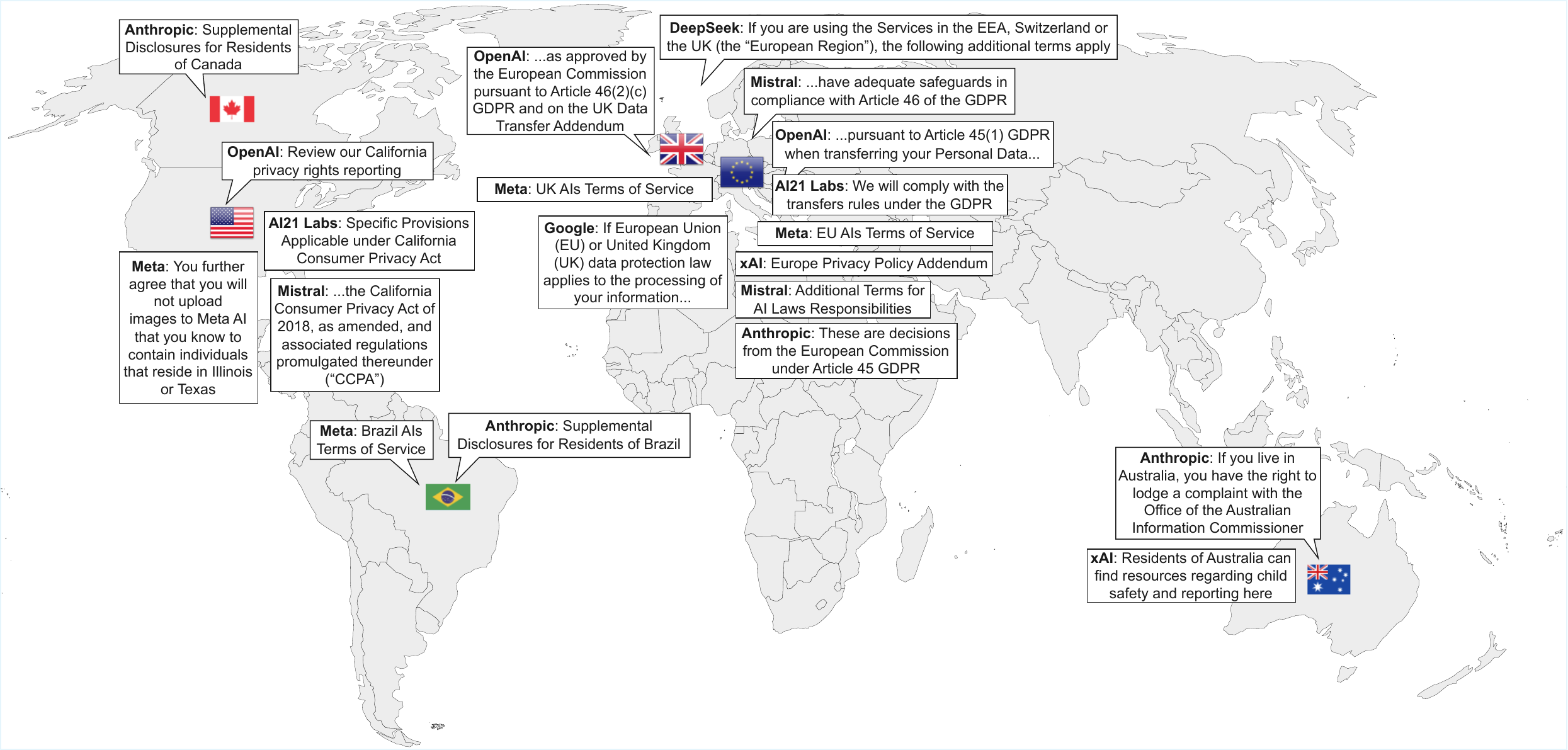}
  \caption{Regulations and regions mentioned in LLM privacy policies and supplementary documents that providers most visibly intend to comply with. The courtesy of the world map is to~\cite{BlankMap}. 
  % \chen{may use the map that is white inside rather than gery which makes it a bit hard to read.}
  }
  \label{fig_geo_regu}
\end{figure*}
%-----------------

% \zhen{Jurisdictions (disparities in geographical regulations)}
% \zhen{What are the regulatory ``hot spots'' referenced by LLM providers? such as EU, california...what actions taken by the LLM providers in these regions, such as EU-targeted sections (data transfer and user rights), Jurisdiction-specific versions, California rights notices, separate supplementary documents. GDPR and AI Act presence. World map + examples.}
% \zhen{1. hot spots and notice forms; 2. In each region...regulation, what article, what topics; 3. trend, specific to general?}

Across LLM providers, geographic references concentrate in a few jurisdictions that create binding obligations with materially higher compliance risk. Among these jurisdictions, the European Union (EU) is a hot spot and dominated by the GDPR for lawful processing, cross-border transfers, and data subject rights. U.S. state privacy regimes follow behind, with the CCPA being the most prominent regulation mentioned. LLM providers also surface UK, Brazil, Canada and Australia. We present the relevant quotas from privacy policies and supplementary documents in Figure~\ref{fig_geo_regu}. We observe LLM providers respond to regulatory requirements in specific jurisdictions in 3 forms: (1) jurisdiction-specific privacy policy versions, such as OpenAI's Europe Privacy Policy~\cite{OpenAIEUPP}; (2) region-targeted sections in privacy policies, such as \textit{Supplemental Disclosures for Residents of Brazil} in Anthropic's privacy policy; and (3) separate supplemental documents, such as Meta's \textit{AIs Terms of Service (UK)}~\cite{MetaAIsTermsUK}. To understand the impact of geographic regulations on LLM privacy policies, in this section, we discuss how regulations are respected in policy content.

\textbf{European Union.} LLM providers frequently anchor disclosures on data transfers and legal bases in the GDPR, with several making article-level references. OpenAI and Anthropic refrt to Art. 45 and 46 on cross-border data transfers. Mistral claims to guarantee service providers have adequate safeguards under Art. 46 when processing personal data outside the EU, and cites Art. 32, ``\textit{Security of Processing}'', in obligations of Mistral AI in its \textit{Data Processing Addendum}. Examining the usage of terms specific to the GDPR is one common approach to study GDPR's impact on privacy policies~\cite{belcheva2023understanding, amos2021privacy, degeling2018we}. In our dataset, five providers explicitly identify the ``data controller'', a term defined in the GDPR, including EU-based Mistral and China-based DeepSeek. Policies commonly list legal bases, e.g., legal obligation and legitimate interests, under Art. 6. Other GDPR-driven elements, such as data subject rights and choices, appear without article citations, suggesting that GDPR’s practical impact exceeds explicit references. In addition to GDPR, the EU AI Act has prompted Mistral to publish \textit{Additional Terms for AI Laws Responsibilities}, clarifying parties’ obligations under the Act.

% \textbf{European Union.} LLM providers often reference the GDPR in their disclosures regarding data transfers and legal bases. Several providers show explicit article-level references. Specifically, OpenAI and Anthropic refer to GDPR Article 45 and 46, which govern the cross-border transfers of personal data. Mistral claims to guarantee service providers have adequate safeguards in compliance with GDPR Article 46 when processing personal data outside the EU, and mentions Article 32, ``\textit{Security of Processing}'', in one of the obligations of Mistral AI in its \textit{Data Processing Addendum}. Examining the usage of terms specific to the GDPR is one common approach to study GDPR's impact on privacy policies~\cite{belcheva2023understanding, amos2021privacy, degeling2018we}. We observe that 5 LLM providers explicitly identify the ``data controller'', a term defined in the GDPR, including not only providers from the EU, e.g., Mistral, but also DeepSeek from China. Legal bases for processing personal data, such as legal obligations and legitimate interests, are often included in policies, as required by GDPR Article 6. GDPR-driven contents also include disclosure of data subject rights and choices that described without GDPR article citations, we therefore believe that GDPR’s practical impact is broader than the explicit references suggest. In addition, the EU AI Act prompted Mistral to prepare an \textit{Additional Terms for AI Laws Responsibilities}, which outlines each parties’ obligations under this Act.

\textbf{United States.} The California Consumer Privacy Act (CCPA)~\cite{CCPA} is the most frequently referenced U.S. regulation. Mistral’s \textit{Data Processing Addendum} includes a dedicated CCPA subsection and commits not to ``sell'' or ``share'' personal data, adopting CCPA terminology. AI21 provides a long section on specific provisions applicable under CCPA, which lists categories, sources, purposes, and consumer rights. This reflects a common pattern that specific terms are hard-coded into the policies and California-specific transparency is considered as a baseline in the U.S. regimes. Moreover, Meta explicitly names Illinois and Texas in \textit{AIs Terms of Service}~\cite{MetaAIsTerms}, requiring users not to upload images containing residents of those states unless legally authorized to consent. We consider this as a response to Illinois’ Biometric Information Privacy Act (BIPA)~\cite{BIPA} and Texas’ Capture or Use of Biometric Identifier Act (CUBI)~\cite{CUBI}, which regulate collection and use of biometric identifiers for a commercial purpose. This statement echoes Facebook’s 2011 tagging feature controversy that ultimately deemed to violate BIPA and resulting in a \$1.4B settlement~\cite{MetaBIPA}. Compared to the CCPA sections, this constraint is highly localized and modality-specific. Overall, CCPA still functions as the U.S. omnibus template for consumer rights and transparency, while state laws with strong enforcement also have influence, but are limited to specific targets, as Meta is the only LLM provider that mentions Illinois.

\begin{figure*}[t]
  \centering
  \includegraphics[width=.95\linewidth]{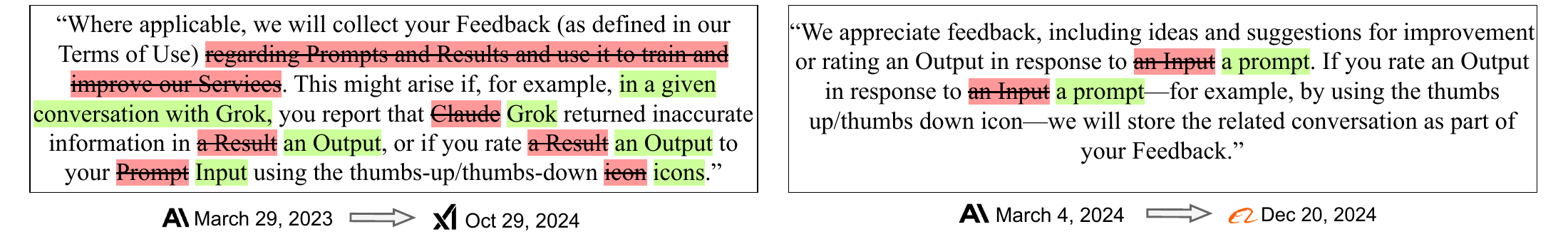}
  \caption{Copy-paste-modify pattern among LLM privacy policies. We observe that xAI and Alibaba may have borrowed from Anthropic's privacy policy when describing the new information type ``\textit{Feedback}'' (thumbs-up/thumbs-down behaviors). Texts are highly overlapping, and only small modifications occur.
  }
  \label{fig_copy}
\end{figure*}
%-----------------

\textbf{Brazil, Canada, and Australia.} In addition to EU and U.S., regulations in other regions are also reflected in the privacy policies. For Brazil, Anthropic dedicates a Brazilian General Data Protection Law (LGPD)~\cite{LGPD} section that defines lawful bases, spells out data subject rights, and details international transfer safeguards relying on standard contractual clauses (SCCs) approved by the Brazilian Data Protection Authority. Meta provides a separate \textit{Brazil Meta AIs Terms of Service} that adapts governing law and jurisdiction for Brazilian users. In Canada, Anthropic’s supplemental disclosures emphasize consent-centric model and cross-border transfer transparency. In Australia, xAI surfaces the jurisdiction in a child safety context, directing residents of Australia to the reporting mechanisms under Australia’s Online Safety Act~\cite{OnlineSafetyAct} for serious online abuse. Overall, these regions illustrate different focuses, showing that LLM providers tune their privacy text to the specific pressure points of different regimes.

In addition, we observe a sign that some LLM providers tend to gradually de-specify regulations, shifting from naming particular laws to using broader phrases such as ``applicable laws''. For instance, OpenAI changed the section title from ``California privacy rights'' to ``Additional U.S. State Disclosures'' and replaced explicit ``CCPA'' mentions with ``local law'' in the body text, so that the same section can cover multiple state regimes without constant renaming or rewriting. Similarly, Mistral folds GDPR and CCPA into a general definition of ``Applicable Data Protection Law''. 
% We also discuss this in Section~\ref{RQ3_content_evo}. 
These signs suggest LLM providers aim to accommodate the growing complexity and divergence of regional regulations within a reusable compliance narrative by reducing the granularity of explicit references.

\subsection{Comparison between LLM Providers.}

\textbf{Policy Coverage.} No single LLM provider’s privacy policy achieves full coverage. Some fine-grained attributes, such as specific ``\textit{Information Type}'' and ``\textit{User Type}'' in certain categories are never or almost never mentioned. At the company level, Anthropic, AI21 Labs, and Mistral cover the most taxonomy attributes, while OpenAI and xAI follow closely and other providers relatively cover fewer attributes. Common topics, such as high-level First-party Data Collection and User Choice \textit{scope}, are widely mentioned and indicate a shared baseline of what LLM providers deem essential to disclose. In contrast, Anthropic explicitly cover nuanced aspects such as policy change type and detailed user type distinctions, which are rarely mentioned by other LLM providers. Nuanced dimensions are more patchy, suggesting that mainstream LLM policies commonly cover high-level compliance topics but diverge once the content zooms into more granular distinctions.
Results in Table~\ref{tab_taxonomy_short} show that U.S. providers (P1–P3) cover 69.0\% of listed attributes on average, while Chinese providers (P5–P7) cover 59.8\% items in the taxonomy. 
Figure~\ref{fig_geo_regu} suggests that U.S. providers more often include region-specific clauses. 
This difference might be caused by that U.S. providers market more broadly across jurisdictions, and therefore their privacy policies need to cover more cross-border and data rights topics; whereas Chinese providers appear more domestically oriented.

\textbf{Copy-paste-modify Pattern.} 
% \chen{May briefly talk about the process, steps?} 
% We observe cases that some borrow external templates or peer clauses, transplanting it into own policy and performing edits.
While reviewing privacy policies, we observe suspicious signs of such as copy-paste-modify, suggesting that the borrowing of policy clauses, a phenomenon revealed in recent studies~\cite{nenadic2025automated, pan2024trap}, is also seen in mainstream LLM providers. 
They might borrow external templates or peer clauses, transplant them into own policies, and perform basic edits.
Two examples are shown in Figure~\ref{fig_copy}. Anthropic’s clause about collecting ``Feedback'' regarding prompts and results, which is a new data type, appears to be a template that xAI and Alibaba have adapted with light edits. Modifications we observe include (1) lexical-level edits, such as ``prompt'' with ``input'' and enriching sentences with extra terms; and (2) sentence-level modifications, such as deleting model training as the purpose for data collection. 
We find that similar clauses appeared in descriptions of data collection and processing. 
This may be because companies, facing new data interaction in LLM scenarios, chose to borrow from peers to prevent significant oversights, especially in the early stages. 
Also, prior work shows that companies may lean on templates to cover mandatory elements~\cite{pan2024trap}.
% \chen{any reference to it? Any other reasons, may ask ChatGPT :-) Does the previous works on privacy policy for general software also share similar copy-paste issue or not?} 
Legal teams may start from existing ``good'' language from competitors’, then make changes to reflect internal differences. This can have legitimate motivations, as converging on a common vocabulary may help regulators and auditors compare practices and reduce drafting costs. However, boilerplate languages creates potential issues for transparency and accountability that policy text can be only partially adapted, leaving residual references that do not match actual product behavior~\cite{sun2020quality}. 
\section{Longitudinal Analysis of Evolution of Privacy Policies for LLMs (RQ3)}
\label{sec_rq3}

 %-----------------
\begin{figure*}[t]
  \centering
  \includegraphics[width=.96\linewidth]{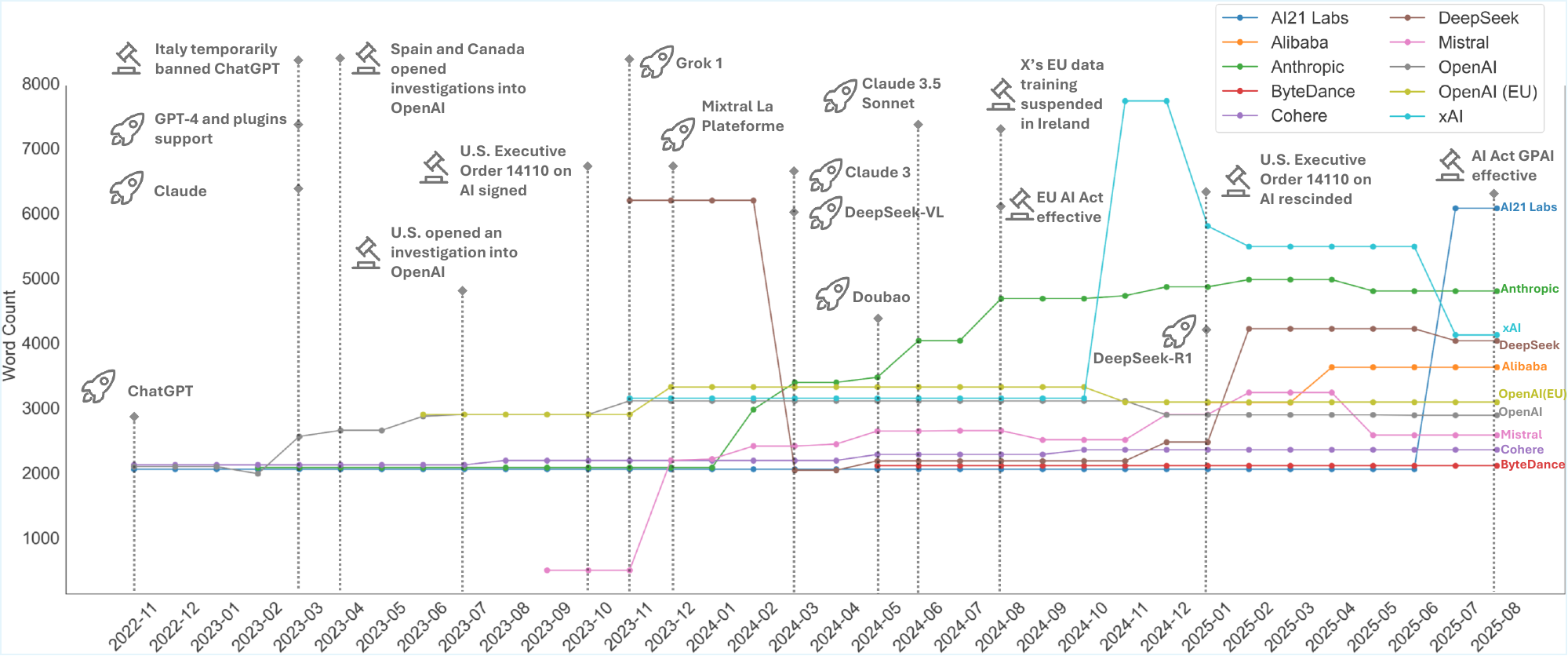}
  \caption{Timeline of policy changes. Key events include product releases (\icon{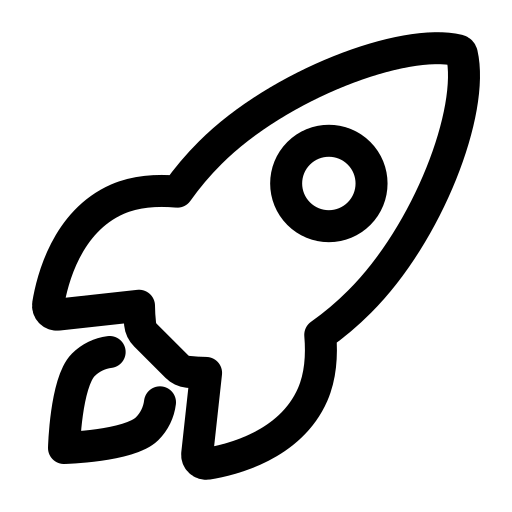}) and regulatory actions (\icon{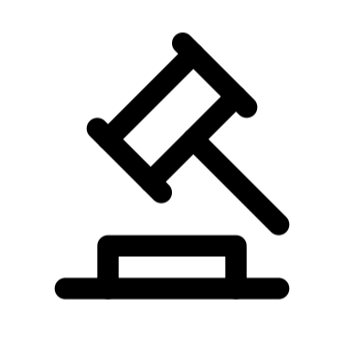}).
  % \chen{1. may add number for each event for easier reference? 2. Too many colors in the figure, hard to tell which color is for which company. May put the company name/icon legend directly to the last column of the line? Or any other better way?}
  }
  \label{fig_word_count_line}
\end{figure*}
%-----------------

% How Class Action Lawsuits Shape Data Privacy Rights and Obligations~\cite{fenske2023courting}.
To examine how privacy policies evolve and whether key ecosystem events may affect those changes, we compile a timeline of product and feature releases and regulatory actions for mainstream LLM providers and aligned each company’s policy updates to this timeline. We then apply a sentence-level heuristic between consecutive policy versions to detect edits and categorize them based on our taxonomy. The details of this methodology is described in Section~\ref{PP_analysis_method}, and the results are presented in this Section.

% \zhen{Focus on the topics extracted in RQ1 and RQ2 and observe how these points evolve, such as the description of data training (Italy). Find events based on points and try to avoid Joint impact.}

% \subsection{Update Frequency}
% \label{rq3_update_frequency}

% Update Frequency doesn't match version count, reasons:
% 1. Adjusted the capitalization and abbreviation of the company name (xAI )

% Interesting examples of changes:
% If you include personal information in your Prompts, we will also process that information.	
% If you include personal data in your Prompts, we will collect that information and this information may be reproduced in the Outputs.

%You may have the right to request a copy of the personal information Anthropic processes about you, subject to certain exceptions.	
%You may have the right to request a copy of the personal data Anthropic processes about you, subject to certain exceptions.

\subsection{Policy-change Drivers}
\label{rq3_driver_sec}
% \zhen{What are the primary categories of policy-change drivers? What useful information can various stakeholders gain from viewing the graph? For example, a lawmaker might find that some event nodes haven't changed significantly...More holistic, why other companies didn't change?}
% Temporal alignment rather than strict causality.
% Regulatory pressure and product release are the primary policy-change drivers.
% Lag and synchronization.

% , regulatory changes (\icon{Figures/icons/8581254_auction_court_law_sale_icon_big.png}), and corporate developments (\icon{Figures/icons/11210199_kim_hands_finance_agreement_sign_icon.png}

We collect two categories of events in the LLM ecosystem:
(1) \icon{Figures/icons/9035573_rocket_outline_icon.png} Product and feature releases, such as the launch of new services and new models; 
% is a common trigger for visible policy growth. 
(2) \icon{Figures/icons/8581254_auction_court_law_sale_icon_big.png} Regulatory actions, including enforcement actions taken by data protection authorities; 
% also impact privacy policy updates. 
% (3) \icon{Figures/icons/11210199_kim_hands_finance_agreement_sign_icon.png} Corporate developments, such as establishment of partnerships. 
% show weak temporal alignment with policy edits. 
We align policy captures with events in Figure~\ref{fig_word_count_line}. Note that the inflection points fall at the timestamps of two privacy policy captures and are not strictly consistent with the ``effective date'' in the policy text. We emphasize that our goal is not to draw strict causality but to show temporal alignment, and we interpret these events as correlations rather than proof of cause.

\textbf{Product and Feature Releases.} In multiple cases, policy updates occur in step with feature launches. For instance, The earliest xAI privacy policy we captured was from November 2023, coinciding with the release of Grok 1. Also starting this month, as different models are gradually released and open-sourced, DeepSeek has synchronously overhauled its privacy policy with broader service coverage and international wording for international growth. Aligning with the product launch, Mistral’s privacy policy was significantly updated in December 2023, changing from ``Mistral AI does not collect private or customer data'' to data collection disclosures. These product-aligned jumps tend to concentrate changes in data collection, disclosure and user controls, consistent with new services that requires fresh disclosure. We observe that policy updates affected by product launches are often synchronized, reflecting the material preparation before the launch of new services.

\textbf{Regulatory Actions.} Some policy edits align with regulatory events. OpenAI extended its policy with lawful-basis explanations and notices on model training and children consent one month after being banned in Italy in March 2023 for allegedly violating transparency provisions of the GDPR~\cite{gdprhubItaly}. Similarly, the Irish Data Protection Commission launched proceedings in Ireland's High Court on X's use of EU user personal data to train Grok on Augest 2024~\cite{IrishDPCpressure}. xAI added dedicated Europe section with data controller details and lawful-basis breakdown nearly 3 months later. We observe that while product releases tend to precede policy edits synchronously, regulatory milestones often show longer lags. This is expected, as policy teams typically push updates only after engineering and legal due-diligence finish, and interpreting guidance and coordinating with data protection officers (DPOs) also takes time. In addition, certain regulatory events show no visible impact. For instance, U.S. Executive Order 14110~\cite{EXECUTIVEORDER14110} shows little synchronized spike for U.S. providers. This may be related to the nature of specific regulation, e.g. EO 14110 mostly targets federal agencies and standards rather than consumer privacy notices.

% \textbf{Corporate Developments.} Corporate events can spark public discussion and privacy concerns~\cite{ali2025understanding}, but they rarely align with privacy policy edits. We therefore do not include such event in Figure~\ref{fig_word_count_line}.
% % which are user-facing privacy disclosures. 
% Partnerships between companies may be absorbed by existing service provider clauses, and legally significant changes usually reflect in contracts, not the public notice.

Overall, observable impacts commonly arise from product releases and company-specific regulatory actions. We find that policy edits are company-specific and episodic. Key events tend to impact specific companies, with few events having broad cross-company impacts. Product releases lead to internal policy updates, and regulatory actions usually targets at a specific company rather than across the board. Product and feature releases are common drivers, and regulatory actions also show strong temporal alignment, yet their effects are targeted and often lagged, concentrating on the company behaviors and jurisdictions in scope.

% \sd{there are also word-level adjustment. E.g., ``We use the information we get from Google Analytics to maintain and improve the Site and our products.''	AND ``We use the information we obtain from Google Analytics to maintain and improve the Websites and our products.''}

\subsection{Content Evolution}
\label{RQ3_content_evo}
\begin{figure}[t]
  \centering
  \includegraphics[width=.98\linewidth]{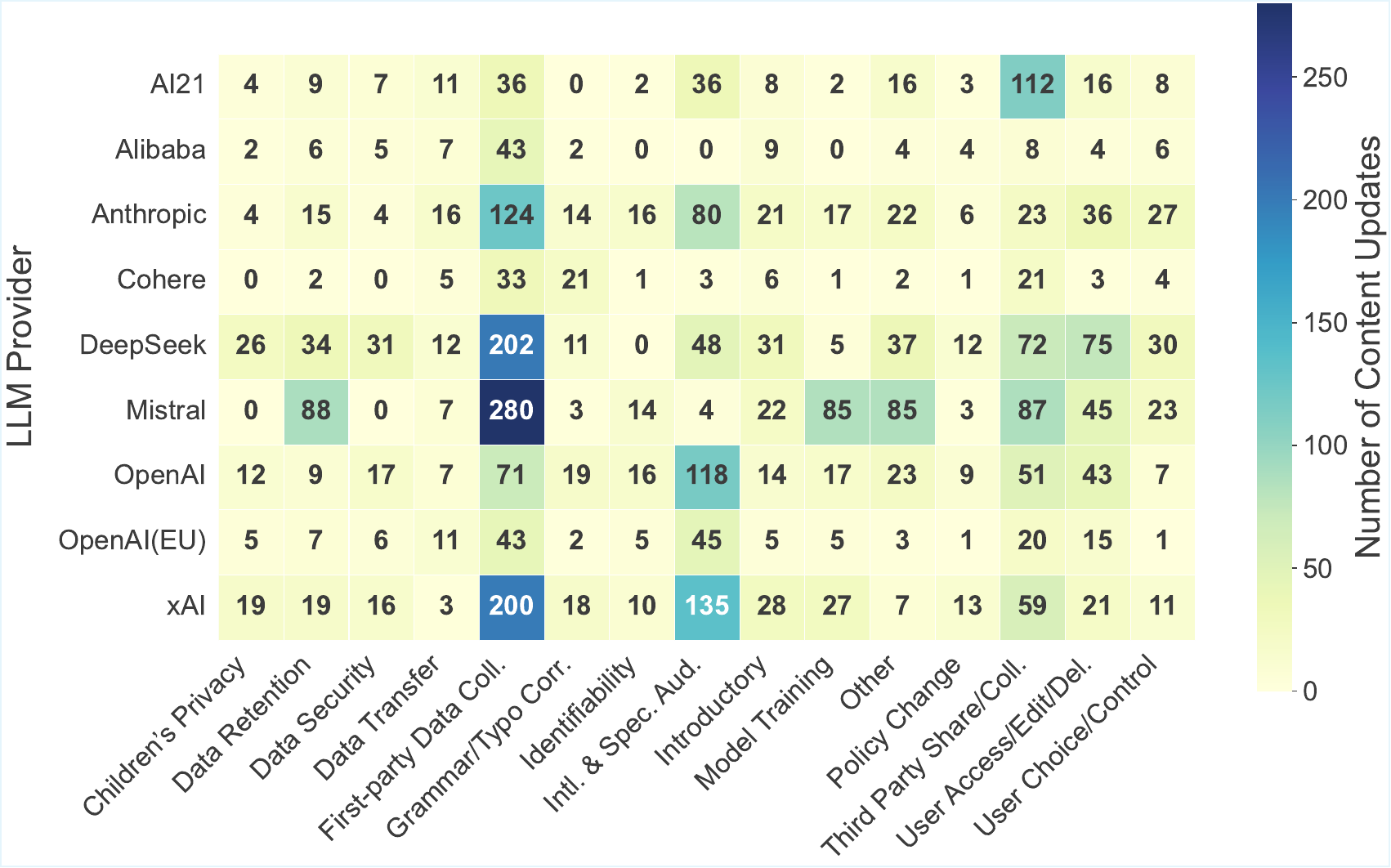}
  \caption{The distribution of policy edits for categories across LLM providers. 
  % \sd{presentation. Enlarge font size and overall visibility.}
  }
  \label{fig_content_change}
\end{figure}

% \chen{Try to align the figur with the content in the same page.}
We discuss the content evolution in terms of categories. The distribution of policy edits is presented in Figure~\ref{fig_content_change}. The most edits fall on First-party Data Collection, and International and Specific Audiences also shows frequent edits. This highly uneven distribution illustrates that LLM providers have a clear tendency to update the policy content.

\textbf{First-party Data Collection.}
Most providers made the most policy edits on the First-party Data Collection. This section often lists the data type and how it was collected, purposes, and how the data will be used. 
% As discussed in Section~\ref{rq1_patterns}, providers tend to numerate data sources in different groups. 
Policy edits include changes to data types and their interpretations. For instance, Mistral removed the sentence \textit{``if you include personal data in Your Input, then such personal data may be included in the Output''} from explanations of data type \textit{``Output''} in the version effective 27 May 2025. On March 14, 2023, OpenAI added a new collected data type \textit{``User Content''}, such as personal information included in the input and file uploads. 
% \chen{Do not understand how it is related to first-party data collect.} 
This edit may be related to the release of ChatGPT at the end of 2022, as this policy version is the first official update since the release. Frequent changes in interpretations may prevent users from forming a consistent understanding of the data being collected and its possible consequences.

\textbf{Third-party Products/Services.}
Edit in third-party content is prominent for AI21 Labs, and is also notable for Mistral. The changes often occur in explanations of third-party entities and relevant interaction scenarios. For instance, in version captured on Dec. 14, 2024, Mistral added two new third-party providers, Black Forest Labs~\cite{BlackForestLabs} (image-generation) and Brave, Inc.~\cite{Bravebrowser} (search-engine), with both accessing user input and the first one also accessing the output. This is consistent with LLM gradually expanding features through third parties, which creates new outbound data flows. Disclosure of data interactions can be vague. When stating research as a purpose of data usage, OpenAI removed the sentence \textit{``which may remain internal or may be shared with third parties, published or made generally available''} in version updated March 14, 2023. On Sep. 19, 2022, OpenAI listed \textit{``other users''} as a third party to receive personal information and added the sentence \textit{``certain actions you take may be visible to other users of the Service''}. This recipient and its confusing description were removed in the next policy version, and eventually returned with specific examples, including custom actions for GPTs and sharing ChatGPT conversations. 
% Insufficient disclosure, especially in the context of LLM interaction scenarios, is noteworthy. \chen{Is this last-sentence summary the common norm in security paper? Or put them into discussion.}

 %-----------------
%-----------------

\textbf{Identifiability.}  
LLM providers often emphasize the user data used for model training has been aggregated or de-identified, and we observe changes in tone. OpenAI once removed the \textit{``may''} from \textit{``we may aggregate...''} in the EU policy version to affirm its actions, but added the word back in the latest version.
Certain claims could be ambiguous. Mistral once stated \textit{``we use \underline{commercially reasonable efforts} to de-identify such data before using it to train or improve our models.''} and deleted it on 27 May 2025. Moreover, this process is not irreversible. In policy updates, OpenAI and Authropic added that they may re-identify information if necessary or required by law.

\textbf{Model Training.} The scope of data usage for model training is expanding. LLM providers often added \textit{``publicly available information on the Internet''} as a data source. In addition, the concept of \textit{``users''} and \textit{``non-users''} has been gradually implied in data processing and opt-out choices. When disclosing data from third-party sources used for model training, Anthropic explicitly stated that personal data of \textit{``non-users''} would also be collected and processed (effective February 19, 2025). 
% DeepSeek added \textit{``publicly available information via the Internet sources''} as data for model training (updated February 14, 2025). 
 Mistral added a qualifier, ``on your Mistral AI Account'', to the original sentence to limit the enforcement to users with an account. Under regulatory pressure, disclosure regarding model training has been improved. For instance, after Italy's temporary ban on ChatGPT in 2023~\cite{gdprhubItaly}, OpenAI added supplemental documents for opt-out and model development. However, privacy protection through the training process still lacks detail and wording varies. Anthropic added \textit{``we take steps to minimize the privacy impact on individuals''} (Effective March 4, 2024), while Mistral added \textit{``we take steps to make sure that Your Personal Data remains protected and secure throughout the training of Our Model''} but removed in the next policy version. Mistral also claimed that it is \textit{``likely''} that personal data will be used for model training.
 % even if \textit{``best practices''} are used to filter them (downgraded to \textit{``good practices''} in later versions).  \chen{It was mentioned in the previous part of the paper. Overlap here.}

\textbf{International and Specific Audiences.}
This category includes additional information about how providers disclose personal information of individuals in specific regions, such as the European Economic Area and California. OpenAI has made the most policy edits, changing additional details on user data and additional user rights, such as California privacy rights. For instance, in its March 14, 2023 update, OpenAI removed the description regarding data sources and usage, and added the right to correct the personal information for California residents. In subsequent updates, OpenAI changed the section title from \textit{``California privacy rights''} to \textit{``Additional U.S. State Disclosures''} and replaced CCPA~\cite{CCPA} with \textit{``local law''} in the paragraphs to generally cover laws in more regions. Similarly, xAI renamed \textit{``Additional U.S. state disclosures''} to \textit{``Privacy rights and choices''} and replaced \textit{``U.S. privacy laws''} with \textit{``applicable privacy laws''} in the paragraphs. Children's privacy is an important topic. We observe variations in the age gate for children. xAI once considered persons under 18 years old as children, but later lowered the limit to 13 years old. DeepSeek used to regard 14 years old as children, but later not specifically mention the age gate. Such changes may also be influenced by regulations in different regions.

\textbf{Content Migration.} Over time, certain portions of the privacy policy may be split into supplemental documents. For instance, we observe jurisdiction-specific supplements are provided. In xAI, privacy rights under European privacy laws were moved to \textit{``Europe Privacy Policy Addendum''} on February 14, 2025. This modularization may reduce change risk, as providers can revise one regional supplemental document without rewriting the core policy. Mistral moves the content regarding third-party service providers to its Trust Center, moving cookie-related contents to \textit{``Cookie Policy''}, and move definitions section to \textit{``Data Processing Addendum''}. Moving content to supplemental documents is becoming a trend, and we discuss it in Section~\ref{dis_sec_layer}.
% \chen{So more split or merging? any merging?}

% xAI: ``18'' (Updated: November 11, 2024) to ``13'' (Updated: December 20, 2024).
% xAI: \textit{``teenagers between the ages of 13 and 18 years old''} (Effective: January 2, 2025) to \textit{``teenagers between the ages of 13 and 17 years old''} (Effective: February 14, 2025).
% DeepSeek: ``14'' to not specifically mention the age gate (Last Update: July 4, 2025).

% \textbf{Wording, Grammar Consistency and Typo Correction.} 
\textbf{Terminology Harmonization.}
Policy edits also fall into terms wording and service scope to enhance regulatory alignment and operational precision.
For instance, ``Social Media Pages'' was modified as ``social media pages/activity'', lowering capitalization and shifts from page‐level to activity‐level framing. xAI changed the word ``Service'' to ``Grok'' but then changed back to ``Service''. Such edits shows oscillation between narrow and generic scopes. These edits improve lexical consistency and also subtly widen scope. 
OpenAI added a Word ``only'' into a sentence ``\textit{...store Personal Information \underline{only} in the course of...}''. 
% describing when the personal information will be processed. 
This word is legally meaningful and is a substantive tightening that places a clear proportionality constraint on processors.  
% \chen{This is very sudden, why? any detailed example?}
In addition, multiple providers changed ``\textit{personal information}'' to ``\textit{personal data}'', aligning the wording with GDPR. Edits on terminology can re-draw the boundaries of data practices or even quietly re-scope data flows.

\section{Discussion}
\label{sec_discussion}
% \zhen{more connected to privacy policies. Shorter. in one page. Structure: refer to our findings, compare with relevant work, do they align? If not, what should we do...}

\subsection{User Privacy Expectations in LLM Services}
% User privacy expectations:
% LLM providers’ trade-off: legal requirement/harmful to user trust
% The contradiction of user needs: wanting to know details, but also wanting to be easy to understand (paradox)
% User’s willingness~\cite{balash2024would}.
% Privacy nutrition labels for open-source generative AI-based applications~\cite{si2024solution}.

% Data interaction scenarios in LLM services differ fundamentally from general software, reshaping user privacy expectations. 
% Instead of static profile fields and device permissions, LLM providers routinely process free-form prompts and data flowing through external plugins, often for additional purposes such as model training and safety evaluation.
In Section~\ref{rq1_patterns}, we show that LLM data practices differ from those of general software, which may reshape user privacy expectations.
Users face a long-standing paradox that they expect trustworthy detail yet also concise, digestible explanations. In app context, such tension and related misconceptions hinder informed choices~\cite{cao2021large, wumodeling}. Richer privacy labels for iOS apps can heighten risk perception and reduce willingness to install without resolving understanding gaps~\cite{balash2024would}, suggesting that ``More detail'' alone does not guarantee better decisions. Users often continue using data-intensive services and treating opaque data practices as an unavoidable cost of participation in digital life~\cite{draper2019corporate}. These issues persist and intensify in the LLM context, where conversational agents feel more intimate and human-like, users often misunderstand key policy concepts, and may not anticipate how their data can be retained or inferred over time. For instance, 
long-term memory can surface old inputs.
% in systems with long-term memory, information entered months earlier may unexpectedly resurface in a new conversation and surprise users. 
LLM privacy policies must move beyond simply being longer, toward transparency mechanisms that better align with users’ mental models and cognitive constraints.
% Addressing this expectation gap requires transparency mechanisms that reconcile users’ desire with their cognitive and time constraints, rather than assuming that longer, more detailed policies alone can restore meaningful control.

\subsection{The Layered Paradigm of Privacy Documents} 
\label{dis_sec_layer}
% Our results show that 
LLM privacy policies form a layered pattern. Main policies retain the traditional backbone and add LLM-specific overlays, while pushing fine-grained details into supplemental documents, e.g., Mistral moved the ``Definition'' of terms from its privacy policy to \textit{Data Processing Agreement}. In our dataset, the common supplemental themes are about data processing, product usage, Data deletion and model training. OpenAI, Mistral and Anthropic all prepare a~\textit{Data Processing Addendum/Agreement}. This suggests that supplemental documents are often used to clarify data lifecycle operations and carve out LLM-specific practices. LLM providers increasingly rely on modular documentation to satisfy heterogeneous legal and customer requirements while keeping the main policy relatively stable. 
% This layered ecosystem brings advantages but also drawbacks. Offloading detailed topics to dedicated documents may improve completeness and maintainability.
% However, distributing information across various documents can exacerbate cognitive burden and raises fragmentation and findability concerns that users and even policymakers may struggle with information hunting. 
However, scattering information across documents increases cognitive burden and findability issues.
Prior work on policy formats showed that users appeared not go beyond the initial layer~\cite{mcdonald2009comparative}. Recognizing this problem, OpenAI and Anthropic have enabled document inquiry feature to help users retrieve and understand content. Work on contextual privacy policies~\cite{windl2022automating, pan2024new, chen2025clear} also suggests a promising direction to provide concise, context-aware explanations into the interaction flow with end users.
% \chen{May draw some lessons from privacy policy for general software e.g., explicit data access like location, contact which can be switched off easily. May also use your previous work to make the privacy policy more interactive with end users...}
% data processing, e.g., OpenAI's ~\textit{}, product usage, e.g., Google's ~\textit{}, Data deletion, e.g., Anthropic's~\textit{}, and model training, e.g., OpenAI's~\textit{}.

\subsection{The Future of AI Regulation} 
The effect of GDPR~\cite{GDPR} illustrates how a regulation can profoundly impact privacy policies worldwide. Adopted in 2016, GDPR’s transparency principles have driven measurable changes in privacy policies, such as data retention, legal basis for data processing and user access rights~\cite{linden2018privacy, belcheva2023understanding}. 
These findings are consistent with the Brussels Effect~\cite{bradford2020brussels}, in which companies aligned their policies with EU requirements, even when not formally obligated to. Our analysis suggests this legacy persists in the LLM era, as LLM providers still follow GDPR terminology in their privacy policies. The EU AI Act~\cite{AIAct}, while the first comprehensive AI regulation with obligations for general-purpose AI (GPAI) such as LLMs, has so far left only faint traces in LLM privacy documents.
% The EU AI Act~\cite{AIAct} is expected to play a similar role for AI-specific governance. It is the first comprehensive AI regulation, introducing a risk-based framework and dedicated obligations for general-purpose AI (GPAI) models such as LLMs, including requirements on technical documentation, transparency and model training disclosures. These obligations could trigger a new Brussels-style externality, where the EU AI Act projects its regulatory model beyond EU and AI companies might standardize their AI disclosures globally. 
We find that only Mistral, an EU company, explicitly references the Act in a supplemental document, while other jurisdictions often pursue divergent AI approaches~\cite{delayairegulation, brusselsmirage}.
A recent legal commentary~\cite{brusselsmirage} characterizes such phenomenon as a ``Brussels Mirage'', and few emulation of the EU regime further dilute the likelihood of a global template. Effective LLM governance will depend on how AI-specific rules can be woven into existing privacy policy frameworks, so that AI regulation complements rather than fragments established privacy disclosures.

% Effective governance of LLMs requires combining established privacy laws and AI regulation coherently. Privacy frameworks such as GDPR continue to govern personal data processing, while AI regulations such as the EU AI Act add layers focused on AI risks, documentation, and technical transparency. This creates both opportunities and challenges. The key task for regulators is to coordinate guidance so that AI-specific rules complement established privacy obligations, rather than fragment them. Whether the AI regulation will ultimately replicate a global pull depends on how these regimes are integrated in practice and on whether they can deliver both legal certainty and meaningful protections in the fast-moving LLM ecosystem.
\section{Related Work}
Numerous studies have examined how privacy policies for websites and mobile apps have evolved over time. Adhikari et al.~\cite{adhikari2023evolution} conducted a two-decade longitudinal analysis of website privacy policies, tracing changes in composition, readability, and structural features. Amos et al.~\cite{amos2021privacy} curated a million-document dataset of web privacy policies and found that the policy length was doubled and GDPR has a significant historical impact on policy changes. Wagner~\cite{wagner2023privacy} similarly studied website policies from 1996–2021 and showed how they evolved in response to privacy regulations. Belcheva et al.~\cite{belcheva2023understanding} analyzed website privacy policies and indicated an increased proportion of vague statements over time. Meanwhile, empirical work specifically targeting LLM privacy policies is still nascent~\cite{king2025user}. The study closest to ours is King et al.~\cite{king2025user}, who manually analyze privacy policies of six U.S. frontier AI developers. They coded policies based on the CCPA~\cite{CCPA} and pointed out essential information about user privacy is often missing. Their analysis is normatively anchored in a single regulatory framework and focuses on a U.S. AI provider set at a particular point in time. In contrast, we take a predominantly measurement-driven, objective perspective and hold a broader, longitudinal, and taxonomy-driven view. 
% of the emerging landscape.
\section{Conclusion}
\label{sec_conclusion}
LLM Privacy policies are the fundamental mechanism for informing users about data practices in LLM services. This paper presents the first longitudinal study of privacy policies for mainstream LLM providers worldwide, showing that they are longer, harder to read, and still highly vague compared to those for general software formats. 
% while disclosing new LLM-specific practices such as model training and third-party data interaction flows. 
Our analysis reveals systematic differences in coverage across providers and regions, as well as the impact of ecosystem events on policy updates, offering empirical insights for improving privacy governance in the LLM era.
\section*{Acknowledgment}

This research was supported by the Seed Fund Program of Columbia University’s Data Science Institute.

\bibliographystyle{IEEEtranS}
\bibliography{References}

\section{Appendix}
% \subsection{LLM Usage Considerations}
% LLMs were used for editorial purposes in this manuscript, and all outputs were inspected by the authors to ensure accuracy and originality.

\subsection{Information Type in the Taxonomy}
\label{append_taxonomy}
%=========
%
%
% \input{Tables/taxonomy_circle}
%
%
%=========

%=========
%
%
\begin{table}[h]
\caption{The full list of data type in the taxonomy.}
\label{table_data_type}
\centering
\begin{tabular}{|l|l|}
% \toprule
\hline
\textbf{\makecell[l]{Information\\ type in\\ OPP-115~\cite{wilson2016creation}}} & \makecell[l]{Financial, Health, Contact, Location,\\ Personal identifier, User online activities,\\ User profile, Social media data, Survey data,\\ IP address and device IDs, Computer information,\\ Cookies and tracking elements, Demographic, \\Generic personal information} \\
\hline
% \textbf{\makecell[l]{New\\ information\\ type in our\\ taxonomy}} & \makecell[l]{Prompt, Uploaded files, Uploaded images,\\ Audio, Feedback, Output, Inferred information, \\Security information, Interaction data, \\ Generic generated data, Generic user content, \\ Generic publicly available information} \\
\textbf{\makecell[l]{New\\ information\\ type in our\\ taxonomy}} & \makecell[l]{Prompt, Attachments \& Input Artifacts, Feedback,\\ Model Output, Inferred information, \\Security information, Interaction data, \\ Generic generated data, Generic user content, \\ Generic publicly available information} \\
\hline
\end{tabular}
\end{table}
\end{document}